%% file: 2023-09-19_draft.tex
\documentclass[]{article}
\usepackage{setspace} 
\usepackage{amsfonts}
\usepackage{amsmath}
\usepackage{graphicx}
\usepackage{pdfpages}
\usepackage{booktabs}
\usepackage{hyperref}
\usepackage{multirow}
\usepackage{tabularx}
\usepackage{enumitem}
\usepackage{authblk}
\newtheorem{thm}{Theorem}[section]
\newtheorem{lem}[thm]{Lemma}

\newtheorem{prop}[thm]{Proposition}
\title{Pseudo-Bayesian unit level modeling for small area estimation under informative sampling}
\author[1]{Peter A. Gao}
\author[2]{Jon Wakefield}
\affil[1]{Department of Mathematics and Statistics, San Jos\'e State University}
\affil[2]{Department of Statistics and Department of Biostatistics, University of Washington}
\date{}
\begin{document}

\maketitle

\begin{abstract}
	When mapping subnational health and demographic indicators, direct weighted estimators of small area means based on household survey data can be unreliable when data are limited. If survey microdata are available, unit level models can relate individual survey responses to unit level auxiliary covariates and explicitly account for spatial dependence and between area variation using random effects. These models can produce estimators with improved precision, but often neglect to account for the design of the surveys used to collect data. Pseudo-Bayesian approaches incorporate sampling weights to address informative sampling when using such models to conduct population inference but credible sets based on the resulting pseudo-posterior distributions can be poorly calibrated without adjustment. We outline a pseudo-Bayesian strategy for small area estimation that addresses informative sampling and incorporates a post-processing rescaling step that produces credible sets with close to nominal empirical frequentist coverage rates. We compare our approach with existing design-based and model-based estimators using real and simulated data.
\end{abstract}

\section{Introduction}

Producing estimates of health and demographic indicators such as child mortality rates at subnational resolutions is valuable for assessing inequality between regions. The problem of reliably estimating subpopulation quantities based on survey data is commonly called small area estimation. Pfeffermann \cite{pfeffermann_new_2013}, Rao and Molina \cite{rao_small_2015}, and Ghosh \cite{ghosh_small_2020} provide recent reviews of research in small area estimation. Small area estimation methods have been used for subnational mapping of a variety of outcomes, including indicators of poverty \cite{bell_overview_2016, marhuenda_poverty_2017, corral_pull_2020}, health outcomes \cite{utazi_geospatial_2020, hogg_two-stage_2023}, and crop yield \cite{erciulescu_model-based_2019}.

When survey data are limited, direct weighted estimators of small area means such as the Horvitz-Thompson \cite{horvitz1952} or H\'ajek \cite{hajek_discussion_1971} estimators can be imprecise or unreliable. Statistical models can improve estimates by incorporating auxiliary covariate information, explicitly accounting for between area variability using random effects, and leveraging spatial dependence to smooth across nearby areas. When survey microdata are available, individual survey responses can be modeled via unit level models. These models are used to motivate estimators of either finite population means and totals or of superpopulation quantities of interest. 

Unit level models, especially those incorporating spatial random effects, are commonly used for mapping subnational health and demographic indicators in low- and middle-income countries (LMIC). These models often account for spatial variation using spatially continuous Gaussian processes, allowing estimates to be generated at arbitrary resolutions. Such models have been used to map a variety of outcomes including child mortality \cite{golding_mapping_2017}, vaccination rates \cite{utazi_geospatial_2020} and disease prevalence \cite{diggle_model-based_2016}.

Ideally, estimates based on statistical models will be robust to model misspecification. When using unit level models with complex survey data, we consider two types of model misspecification. First, the analyst-chosen model for population responses may be misspecified. Second, model-based approaches to small area estimation often assume that the sampling design is ignorable, meaning that the distribution of sampled responses will be identical to that of non-sampled responses, which may not be the case if the survey design features unequal sampling probabilities or clustering. When the sampling design is not ignorable, it is crucial to account for potential differences between sampled and non-sampled units. One approach is to include all variables used to specify the sampling design as predictors in a model, so that a particular unit's response will be independent of whether it is sampled after conditioning on design variables. If we can specify such a model, we can say that sampling is uninformative with respect to the model. For example, if a particular survey design involves sampling clusters with probability proportional to size, cluster size may be included in the model. 

However, we may only observe a subset of relevant design variables or the functional form of the relationship between the design variables and responses may be unknown. In this paper, we describe sampling as being informative with respect to a model if the model does not apply to both the sampled and non-sampled units. We aim to address informative sampling by leveraging design information such as sampling weights. The sampling weight for an individual unit is defined as the inverse of that unit's probability of inclusion in the sample. Sampling weights are commonly used to compute direct weighted estimators such as the Horvitz-Thompson \cite{horvitz1952} or H\'ajek \cite{hajek_discussion_1971} estimators. Area level models commonly used in small area estimation like the Fay-Herriot model \cite{fay1979} account for the survey design by approximating the sampling distributions of these direct weighted estimators. When estimating unit level models, addressing informative sampling with sampling weights is less straightforward. 

Rao and Molina \cite{rao_small_2015} and Parker et al. \cite{parker_unit_2020} review proposed approaches for incorporating sampling weights when fitting unit level models.. These modifications account for some design features such as unequal sampling probabilities, but may not explicitly address informative sampling and must be extended for use with non-Gaussian response variables.

More generically, for inference using parametric models with complex survey data, pseudo-likelihood methods \cite{binder1983} incorporate sampling weights and can achieve design-consistent estimation of model parameters under certain asymptotic assumptions.  Analogously, pseudo-likelihoods can be used to conduct approximate Bayesian inference using pseudo-posterior distributions \cite{savitsky_bayesian_2016}. These pseudo-likelihood methods have been extended to mixed effects models \cite{pfeffermann1998, rabe-hesketh2006, asparouhov2006}, but frequentist maximum pseudo-likelihood estimators can be sensitive to weight scaling \cite{slud2020, savitsky_pseudo_2022}. Moreover, credible sets based on pseudo-posterior distributions do not generally achieve valid frequentist coverage rates, even asymptotically \cite{han2021, williams2021, leon-novelo_fully_2019}. This body of research has generally focused on estimation of fixed effects, treating random effects as nuisance parameters. In the context of small area estimation, prediction of random effects at the small area level is principally important. Although previous research has applied pseudo-Bayesian approaches for small area estimation \cite{parker_computationally_2022}, the issues of weight scaling and miscalibrated interval estimates have not been explored extensively in the context of small area estimation.

In this article, we outline a strategy for conducting pseudo-Bayesian inference using unit level models. As pseudo-Bayesian credible sets for model parameters may not converge on valid frequentist confidence sets due to dependence between units and informative sampling, we adapt a post-processing method proposed by Williams and Savitsky \cite{williams2021} to rescale our credible sets for small area means. In simulations that we report, the rescaled interval estimates achieve close to nominal empirical coverage rates. We apply our strategy for estimating small area means of both continuous and binary response variables. 

The rest of this article is organized as follows. In Section \ref{s:notation}, we outline our notation and describe the combined model- and design-based inferential framework we use to assess our estimators. Section \ref{s:existing} reviews standard estimation approaches for unit level models using sampling weights and Section \ref{s:methods} details our pseudo-Bayesian approach for generating point and interval estimates of small area means. In Section \ref{s:sims}, we evaluate the performance of our approach in simulation, and in Section \ref{s:applications}, we apply our method to estimate vaccination rates using data from the Demographic and Health Surveys. Finally, in Section \ref{s:dis}, we discuss our method and outline directions for future research.

\section{Background and inferential framework}\label{s:notation}

\subsection{Notation}

Let $U=\{1,\ldots,N\}$ index a finite population of size $N$. For all $j\in U$, we let $y_j$ denote the response value of interest for unit $j$ and $\mathbf{z}_j$ denote a vector of auxiliary variables. We assume $U$ can be partitioned into $m$ disjoint administrative areas, $U=U(1)\cup \cdots\cup U(m)$, where $U(i)$ denotes the $N(i)$ indices corresponding to units in area $i$. Let $S=\{j_1,\ldots, j_n\}\subset U$ denote a random set of $n$ sampled indices, where $S=S_1\cup\cdots\cup S_m$ is the corresponding partition by administrative area. 

We assume a probability sampling scheme where for all $j\in U$, $\pi_j$ denotes the probability that $j\in S$, also called the inclusion probability of unit $j$, which may depend on $\mathbf{z}_j$. For all $j\in U$, we define $\delta_j$ to be the inclusion indicator for unit $j$. In other words, $\delta_j=1$ if $j\in S$ and $\delta_j=0$ otherwise. We let $w_j= 1 / \pi_j$ denote the sampling weight for unit $j$.

Following Rao and Molina \cite{rao_small_2015}, we let $y_{ij}=y_j$ if $j\in U(i)$ and $y_{ij}=0$ otherwise. We define $\delta_{ij}$, $w_{ij}$,  and $\mathbf{z}_{ij}$ analogously. We define $\mathbf{Z}$ to be the matrix of auxiliary variables. The finite population small area means $\overline{\mathbf{Y}} = \{\overline{Y}_1,\ldots, \overline{Y}_m\}$ can be defined such that for each $i$,
$$
\overline{Y}_i= \frac{1}{N(i)}\sum_{j\in U(i)}y_{ij}.
$$

\subsection{Unit level modeling for small area estimation}

Unit level modeling approaches to small area estimation relate individual survey responses $y_{ij}$ to unit-specific auxiliary information and borrow strength from similar or nearby areas when estimating a small area quantity. For continuous responses, Battese, Harter, and Fuller \cite{battese1988} introduced the nested error regression model (also called the basic unit level model by Rao and Molina \cite{rao_small_2015}):

\begin{equation}
y_{ij} = \beta_0+\mathbf{x}_{ij}^T\boldsymbol\beta_1+ u_{i}+\varepsilon_{ij} \label{eq-bhf}
\end{equation}
where $\beta_0$ denotes an intercept term, $\mathbf{x}_{ij}$ denotes observed covariate values, and $\boldsymbol\beta_1=(\beta_{1},\ldots, \beta_{p})$ denotes the corresponding coefficients.  We assume that $\mathbf{x}_{ij}$ corresponds to a subset of the variables included in $\mathbf{z}_{ij}$, allowing for the possibility that not all relevant variables used to design the survey are observed. The area level effects, denoted $u_i\stackrel{iid}{\sim} N(0,\sigma_u^2)$ and $\varepsilon_{ij}\stackrel{iid}{\sim} N(0,\sigma_\varepsilon^2)$, represent random and independent unit level effects. For binary or count responses, analogous linear models wtih appropriate link functions an be used for $y_{ij}$.

Under this model, $\overline{Y}_i=\beta_0+\overline{\mathbf{x}}_i^T\boldsymbol\beta_1 +u_i+\overline{\varepsilon}_i$ where $\overline{\mathbf{x}}_i$ and $\overline{\varepsilon}_i$ denote the area means of $\mathbf{x}_{ij}$ and $\varepsilon_{ij}$, respectively. From a model-based perspective, if we view $\varepsilon_{ij}$ as representing noise or measurement error added to the true quantity of interest for individual $j$, then a more appropriate target estimand is
$$\mu_i= E(\overline{Y}_i\mid \overline{\mathbf{x}}_i, u_i)=\beta_0+\overline{\mathbf{x}}_i^T\boldsymbol\beta_1 +u_i,$$
assuming we know $\overline{\mathbf{x}}_i$ for each area.
Another justification for using $\mu_i$ instead of $\overline{Y}_i$ as the target estimand is that even if we view $y_{ij}$ as being measured without error, by the law of large numbers, $\overline{\varepsilon}_i$ converges in probability to $E(\varepsilon_{ij})=0$ as $N(i)\rightarrow\infty$. As such, it is standard in the small area estimation literature to focus on estimation of $\mu_i$ instead of $\overline{Y}_i$ for the basic unit level model.

\subsection{Interpretation of the basic unit level model}

Practically, treating the area specific intercepts $u_i$ as Gaussian random effects explicitly models variability between areas and shrinks small area mean estimates towards $\beta_0+\overline{\mathbf{x}}_i^T\boldsymbol\beta_1$. Traditionally, the observed values of a random effect such as $u_i$ are viewed as draws from some population, but only population characteristics (averaged over $u_i$), and not the draws themselves, are of interest. Hodges \cite{hodges_richly_2016} calls random effects interpreted in this way ``old-style" to distinguish them from ``new-style" random effects, which represent the entire population of interest or represent draws from some distribution from which additional draws cannot be obtained. Under the model (\ref{eq-bhf}), $u_i$ represent area-specific deviations from the global mean. For small area estimation, we are interested in estimates for a fixed set of $m$ areas, so we interpret the random effects as ``new-style" effects. From this perspective, incorporating area-specific random effects produces a flexible model constrained by the Gaussian assumption on $u_i$, preventing overfitting when data are limited.

When specifying a population level model, however, it may be undesirable to use a model of the form (\ref{eq-bhf}) including random effects. We can rewrite the nested error regression model as follows:
\begin{equation}
y_{ij} = \beta_{0i}+\mathbf{x}_{ij}^T\boldsymbol\beta_1+\varepsilon_{ij} \label{eq-bhf-alt}
\end{equation}
where area specific intercepts $\beta_{0i}=\beta_0+u_i$ are independent $N(\beta_0, \sigma_u^2)$ variables and now $\mu_i=\beta_{0i}+\overline{\mathbf{x}}_i^T\boldsymbol\beta_1$. Instead of treating the $\beta_{0i}$ as draws from a Gaussian distribution, we could view them as fixed area specific intercepts from a frequentist perspective, which would make $\mu_i$ fixed across populations after conditioning on auxiliary variables $\mathbf{X}$. Given sufficient data in each area, it could be sensible to use a model with only fixed effects to avoid shrinkage. From this perspective, $\beta_{0i}$ account for stable population level differences between areas that are not explained by differences in the available predictors $\mathbf{x}_{ij}$. From a Bayesian hierarchical modeling perspective, the difference between using ``fixed" effects versus random effects is less salient: the $\beta_{0i}$ parameters are always treated as random and the shift only involves a change in the prior on $\beta_{0i}$. Under the random effects model, $\sigma_u^2$ is also a random variable while under the fixed effects model, $\sigma_u^2$ would be fixed and typically large.

\subsection{Joint model-and-design based inference}

When conducting inference for model parameters such as $\mu_i$, it is common to assume that sampling is uninformative with respect to the model. In other words, one model is assumed to hold for both the sample and population. In practice, this assumption is difficult to verify, as the model must accurately describe the functional form of the relationship between $\mathbf{x}_{ij}$ and $y_{ij}$.

If the model is misspecified for the sample data, then model-based estimators need to be adjusted. Pfefferman and Sverchkov directly address this by modeling the sample inclusion mechanism \cite{pfeffermann2007}. Another approach is to use the population level model to define ``census" model-based estimators for model parameters that could be computed given complete population data. Traditional design-based sample estimators that utilize sampling weights can subsequently approximate these census estimators. This inferential approach accounts for model-based variability in the census estimators and design-based variability in the sample estimators.

We study point and interval estimators of $\mu_i$ under this combined model- and design-based framework, as developed by Rubin-Bleuer and Kratina \cite{rubin-bleuer_two-phase_2005} and also examined by Savitsky and Williams  \cite{savitsky_bayesian_2016} and Han and Wellner \cite{han2021}. We extend our notation to consider a sequence of sampling designs and populations indexed by $\nu$. Let $U_\nu=\{1,\ldots,N_\nu\}$ index a finite population of size $N_\nu$, where $N_\nu$ increases in $\nu$. Let $\mathcal{S}_\nu$ be the collection of all possible subsets of $U_\nu$.

Let $(\mathcal{Y},\mathcal{B}_{\mathcal{Y}})$ and $(\mathcal{Z},\mathcal{B}_{\mathcal{Z}})$ be measurable spaces for the response and auxiliary variables. Assume $\{(Y_j, \mathbf{Z}_j)\in\mathcal{Y}\times\mathcal{Z}\}_{j=1}^{N_\nu}$ are independent and identically distributed random vectors drawn from a superpopulation model on the probability space $(\Omega,\mathcal{F}, P_{(Y, Z)})\equiv(\mathcal{Y}\times\mathcal{Z},\mathcal{B}_{\mathcal{Y}}\times \mathcal{B}_{\mathcal{Z}}, P_{(Y, Z)})$ where $P_{(Y, Z)}$ denotes the superpopulation measure. We use $P_{0}$ to denote the marginal superpopulation distribution of $Y$.

For cluster and multistage designs, we may wish to consider more complicated dependence structures for the superpopulation model. As an example, Rubin-Bleuer and Kratina outline a two-stage super-population model under which the population is partitioned into primary sampling units (PSU), within which responses and auxiliary variables may be dependent. The above notation may be adapted to reflect this dependence structure where $\{(Y_j, \mathbf{Z}_j)\}_{j=1}^{N_\nu}$ are organized into groups of final-stage sampling units that are independent from one another. In order to simplify the exposition, we continue the discussion treating $\{(Y_j, \mathbf{Z}_j)\}_{j=1}^{N_\nu}$ as independent.

Conditionally on $\mathbf{Z}^{(\nu)}=(\mathbf{Z}_1,\ldots, \mathbf{Z}_{N_\nu})$, we can define a sampling design $P_\nu$, which we view as a probability distribution over the space of possible samples $S\in\mathcal{S}_\nu$. We let $D^{(\nu)}=\{\mathbf{Y}^{(\nu)}, \mathbf{Z}^{(\nu)}, \delta^{(\nu)}, \pi^{(\nu)}\}$ denote the data for the $\nu$th finite population where $\delta^{(\nu)}$ denotes the vector of sample inclusion indicators and $\pi^{(\nu)}$ denotes the vector of sample inclusion probabilities. As outlined by Rubin-Bleuer and Kratina, we can construct a product measurable space $(\mathcal{S}_\nu\times \Omega,\sigma(\mathcal{S}_N)\times \mathcal{F}, \mathbb{P})$ where $\sigma(\mathcal{S}_N)$ is the $\sigma$-algebra generated by $\mathcal{S}_N$ and $\mathbb{P}$ is a combined model- and design-based probability measure for $D^{(\nu)}$. We use $P_{0, \nu}$ to denote the marginal distribution of the observed $Y$, accounting for both model- and design-based randomness.

We seek to understand the asymptotic behavior of our estimators as $\nu\rightarrow 0$ under the combined probability measure $\mathbb{P}$. Similarly, when evaluating estimators, we will generally consider average error metrics taking expectations with respect to $\mathbb{P}$.  Under informative sampling, we consider a combined model-and-design based mean squared error for evaluating point estimators:
$$\mathrm{MSE}(\widehat{\mu}_i)=\mathbb{E}_{\mathbb{P}}[(\widehat{\mu}_i-\mu_i)^2]$$ We are also interested in identifying interval estimates $(\widehat{\mu}_i^{-}, \widehat{\mu}_i^{+})$ such that 
$$\mathbb{P}\left(\mu_i\in(\widehat{\mu}_i^{-}, \widehat{\mu}_i^{+})\right)=1-\alpha$$
for some pre-specified level $\alpha$, where $\mathbb{P}$ indicates the joint probability measure.

Note that $P_0$ the superpopulation law for $Y$  does not need to belong to the model specified by the data analyst. We primarily consider estimators based on nested error regression models of the form specified in Equation (\ref{eq-bhf}). In the Appendix, we discuss the impact of model misspecification, finding that if the estimation model is chosen carefully, model-based estimators can provide reasonable estimates of small area means.

\section{Standard estimation approaches}\label{s:existing}

In this section, we review parameter estimation approaches for the nested error regression model, beginning with approaches which assume ignorability of the sampling design. We proceed to review pseudo-likelihood and pseudo-Bayesian approaches that incorporate sampling weights.

\subsection{Parameter estimation assuming ignorability}

First, we consider the model (\ref{eq-bhf-alt}) treating $\beta_{0i}$ as random, under the assumption that sampling is uninformative with respect to the model. As detailed by Rao and Molina \cite{rao_small_2015}, the frequentist approach proceeds by estimating variance components $\{\widehat{\sigma}_u^2, \widehat{\sigma}_\varepsilon^2\}$ via restricted maximum likelihood or a method of moments. Based on these estimates, the empirical best linear unbiased predictor (EBLUP) $\widehat{\mu}_i^{EBLUP}$ can be computed for all $i$.  Either linearization-based approximation or resampling methods can be used to estimate the MSE of $\widehat{\mu}_i^{EBLUP}$. Prediction intervals can be constructed around the EBLUP based on the asymptotic distribution of $\widehat{\mu}_i^{EBLUP}-\mu_i$.

The model (\ref{eq-bhf})  can be reframed as a Bayesian hierarchical model by placing a known prior distribution $g$ (which may depend on hyperparameters $\tau$) on the parameters $\theta$:
\begin{equation}
\begin{split}
y_{ij}\mid\beta_0, \boldsymbol\beta_1, u_i,\sigma_\varepsilon^2&\sim N(\beta_0+\mathbf{x}_{ij}^T\boldsymbol\beta_1+u_i,\sigma_\varepsilon^2)\\
u_i\mid \sigma_u^2 &\sim N(\mathbf{0},\sigma_u^2)\\
\theta=(\beta_0, \boldsymbol\beta_1,\sigma_u^2, \sigma_\varepsilon^2) &\sim \Pi(\theta).
\end{split}\label{eq-bhm}
\end{equation}
Under this model, our targets are the posterior distributions $p(\mu_i\mid \mathbf{Y}^{(\nu)}, \mathbf{X}^{(\nu)}, \boldsymbol\delta^{(\nu)})$. These posterior distributions are not available in closed form, so sample-based approaches are popular. Using samples from this posterior, where we denote the $k$th sample using $\widehat{\mu}_i^{B(k)}$, we can compute posterior summary statistics and credible intervals for $\mu_i$. 

\subsection{Parameter estimation under informative sampling}

Unless sampling probabilities are constant within areas, the EBLUP for the model (\ref{eq-bhf-alt}) is not design-consistent \cite{rao_small_2015}. To address this, You and Rao \cite{you2002} propose a pseudo-EBLUP method for unequal probability sampling designs that incorporates sampling weights when estimating regression coefficients $\boldsymbol\beta_1$. This approach is not intended to address general informative sampling, though it can do so in many cases. As for the standard EBLUP, the variance components $\sigma_u^2$ and $\sigma_\varepsilon^2$ can be estimated via restricted maximum likelihood. Subsequently the fixed effects parameters are obtained by solving weighted estimating equations.

The resulting parameter estimates are used to predict $u_i$ and compute a pseudo-EBLUP $\widehat{\mu}_i^{psEBLUP}$. The MSE of $\widehat{\mu}_i^{psEBLUP}$ can be estimated via linearization-based approximations \cite{you2002} or resampling \cite{torabi_mean_2010}. 

More generically, given a parametric model, pseudo-likelihood methods incorporate survey weights to construct a sample weighted log-likelihood that approximates the full population log-likelihood \cite{binder1983, skinner1989}. Instead of attempting to incorporate all relevant design features when specifying a model likelihood $p_\theta$,  pseudo-likelihood methods propose a particular superpopulation model of interest that could be fit given full population data. The  pseudo-likelihood is subsequently used to approximate complete population inference for superpopulation parameters using sampling weights. If the weights contain information about informative sampling that cannot otherwise be easily incorporated into a regression model or prediction algorithm, then these approaches may yield estimates with reduced bias.

If the population were fully observed, the census log-likelihood would take the form:
\begin{equation}
\ell_\theta(\mathbf{Y})=\sum_{j=1}^{N_\nu}\log p_\theta(y_{j})\label{eq-cll}
\end{equation}
where $p_\theta$ denotes the likelihood of $y_{ij}$ given parameters $\theta$. The census log-likelihood may be approximated via a sample weighted pseudo-log-likelihood \cite{binder1983}:
\begin{equation}
\ell_\theta^\pi(\mathbf{Y})=\sum_{j=1}^{N_\nu}\frac{\delta_{\nu j}}{\pi_{\nu j}}\log p_\theta(y_{j})
\end{equation}
where $\delta_{\nu j}$ and $\pi_{\nu j}$ denote the inclusion indicator and inclusion probability for unit $j$ for the $\nu$-th population. The census log-likelihood and pseudo-log-likelihood may be used to derive census estimating equations and analogously, sample weighted estimating equations. The weighted estimating equations can be used to derive maximum pseudo-likelihood estimates of $\theta$.  More generally, similar estimating equations can be used to estimate any finite population parameter of interest that can be specified as a solution to a system of census estimating equations, even without some motivating superpopulation model.

The pseudo-log-likelihood implies a pseudo-likelihood of the form 
\begin{equation}
\prod\limits_{j=1}^{N_\nu}p_\theta(y_{j})^{w_{ \nu j}}\label{eq-pl}=\prod\limits_{j=1}^{N_\nu}p_\theta(y_{j})^{\delta_{\nu j}/\pi_{\nu j}}.
\end{equation}
The pseudo-likelihood is not a true likelihood due to the introduction of the weights, but by treating it as such, pseudo-Bayesian inference can be conducted for $\theta$, as introduced by Savitsky and Toth \cite{savitsky_bayesian_2016}. 

If the entire population were observed, the population posterior for $\theta$ could be defined as follows, for all measurable subsets $B\subset \Theta$:
\begin{equation}
\Pi_{\nu}(B\mid \mathbf{Y}^{(\nu)})=\frac{\int_B\prod\limits_{j=1}^{N_\nu}p_\theta(y_{j})\Pi(\theta)d\theta}{\int\prod\limits_{j=1}^{N_\nu}p_\theta(y_{j})\Pi(\theta)d\theta}=\frac{\int_B\exp(N_\nu\mathbb{P}_{\nu}\log p_\theta)\Pi(\theta)d\theta}{\int \exp(N_\nu\mathbb{P}_{\nu}\log p_\theta)\Pi(\theta)d\theta}
\end{equation}
where $\Pi(\theta)$ denotes a prior on the hyperparameters $\theta$ and $\mathbb{P}_{\nu}$ denotes the empirical measure based on the $\nu$-th population:
\begin{equation}
\mathbb{P}_{\nu}(t)=\frac{1}{N_\nu}\sum_{j=1}^{N_\nu}t(Y_j)
\end{equation}
where $t$ denotes a measurable real-valued function. When only a sample of size $n_\nu$ is observed, the population posterior distribution can be approximated by a pseudo-posterior distribution replacing the population likelihood with the pseudo-likelihood:
\begin{align}
\Pi_{\nu}^\pi(B\mid D^{(\nu)})&=\frac{\int_B\prod\limits_{j=1}^{N_\nu}p_\theta(y_{j})^{\delta_{\nu j}/\pi_{\nu j}}\Pi(\theta)d\theta}{\int\prod\limits_{j=1}^{N_\nu}p_\theta(y_j)^{\delta_{\nu j}/\pi_{\nu j}}\Pi(\theta)d\theta}\\
&=\frac{\int_B\exp(N_\nu\mathbb{P}_{\nu}^\pi\log p_\theta)\Pi(\theta)d\theta}{\int \exp(N_\nu\mathbb{P}_{\nu}^\pi\log p_\theta)\Pi(\theta)d\theta}
\end{align}
where $\mathbb{P}_{\nu}^\pi$ is the sample weighted empirical measure for measurable $t$:
\begin{equation}
\mathbb{P}_{\nu}^\pi(t)=\frac{1}{N_\nu}\sum_{j=1}^{N_\nu}\frac{\delta_{\nu j}}{\pi_{\nu j}}t(Y_j).
\end{equation}
Note that $\Pi_{\nu}^\pi$ is not a standard posterior distribution due to the introduction of sampling weights, but is scaled to integrate to one. In this sense, inference based on the pseudo-posterior distribution can be viewed as approximating inference based on the population posterior. As with the unweighted posterior, we can draw samples from the pseudo-posterior for $\theta$ and accordingly obtain estimates of $\mu_i$. 

Intuitively, pseudo-posterior credible sets for a superpopulation parameter can be viewed as approximations of the corresponding population posterior credible sets, which would be based on the full population of size $N_\nu$. In general, credible sets based on pseudo-posterior samples will be too conservative. León-Novelo and Savitsky \cite{leon-novelo_fully_2019} observe undercoverage of the credible sets based on pseudo-posterior samples. Parker et al. \cite{parker_unit_2020} provide an example of pseudo-Bayesian inference applied for small area estimation, but do not explicitly discuss the issue of undercoverage. Various solutions have been proposed for this problem, including rescaling of weights and post-processing of pseudo-posterior samples. 

For pseudo-likelihood based approaches, sampling weights are often rescaled so that the weights $w_{ij}$ sum to the sample size or to the ``effective" sample size, defined as the sample size for a simple random sample achieving the same variance for an estimator as with the existing design \cite{kish_survey_1965}. Rescaling methods have been discussed for a frequentist multilevel model \cite{pfeffermann1998} and in the pseudo-Bayesian setting \cite{savitsky_bayesian_2016}.

\section{Proposed approach}\label{s:methods}

In this section, we describe a general pseudo-Bayesian approach for  small area estimation using unit level models. We apply the post-processing rescaling method described by Williams and Savitsky \cite{williams2021} to correct the coverage of credible sets for $\boldsymbol\mu$ based on pseudo-posterior samples. Although pseudo-Bayesian approaches have previously been adopted for small area estimation, they have generally been applied on an ad hoc basis and do not explicitly address miscalibration of the pseudo-posterior credible sets. We assume a sequence of sampling designs such that as $\nu\rightarrow\infty$, $n(i)\rightarrow\infty$ for all areas $i$. Under this asymptotic framework, as $\nu\rightarrow\infty$, direct estimators of small area means become more reliable. 

Han and Wellner \cite{han2021} and Williams and Savitsky \cite{williams2021} establish results on the asymptotic behavior of the pseudo-posterior distribtion and the pseudo-maximum likelihood estimator (pseudo-MLE) under certain regularity conditions. In particular, they establish Bernstein-von Mises type results for the pseudo-posterior distribution and derive the asymptotic sampling distribution of the pseudo-MLE. Both of these distributions are asymptotically normal and concentrate on $\theta^*$, the parameter vector minimizing the Kullback-Leibler divergence $\theta\mapsto P_0 \log (p_0/p_\theta)$. However, their asymptotic covariances do not agree, so credible intervals based on pseudo-posterior distributions will not generally converge on valid frequentist confidence intervals. Note that neither Han and Wellner nor Williams and Savitsky explicitly addresses misspecification of the superpopulation model but both rely upon results of Kleijn and Van der Vaart \cite{kleijn_bernstein-von-mises_2012}, which establishes a Bernstein-von Mises result for misspecified Bayesian models that illustrates the posterior's concentration on $\theta^*$. In the Appendix, we outline how the results of Han and Wellner, Williams and Savitsky, and Kleijn and Van der Vaart can be adapted for this small area estimation context.

\subsection{Computing a pseudo-posterior}

We describe our approach to pseudo-Bayesian inference for the hierarchical model (\ref{eq-bhm}) before applying our strategy to other models. Under the hierarchical model, our parameters of interest are $\theta=(\beta_0, \boldsymbol\beta_1,\sigma_u^2, \sigma_\varepsilon^2)$ and $\mathbf{u}=(u_1,\ldots, u_m)$, so our goal is to approximate the joint population posterior density:
\begin{equation}
p_\theta(\mathbf{u}\mid D^{(\nu)})\propto \prod\limits_{i=1}^m\prod\limits_{j\in U(i)} p_\theta(y_{ij}\mid u_i)p_\theta(u_i)\Pi(\theta)
\end{equation}
where $p_\theta(y_{ij}\mid u_i)$ denotes the density for response $y_{ij}$ given area effect $u_i$ and parameter vector $\theta$, $p_\theta(u_i)$ denotes the density for the area effect, and $\Pi(\theta)$ denotes the prior. To approximate this population posterior, we use the following sampling-weighted pseudo-posterior density:
\begin{equation}
p^\pi_\theta(\mathbf{u}\mid D^{(\nu)})\propto  \prod\limits_{i=1}^m\prod\limits_{j\in U(i)}p_\theta(y_{ij}\mid u_i)^{\delta_{\nu j}/\pi_{\nu j}}p_\theta(u_i)\Pi(\theta)
\label{sw-posterior}
\end{equation}
We can approximate this density via sampling algorithms or numerical approximation and use this pseudo-posterior to conduct inference for $\mathbf{u}$, $\theta$, and subsequently $\mu_i$.

\subsection{Post-processing adjustment}

Generalized posteriors produced by replacing a standard likelihood in a Bayesian analysis with a pseudo-likelihood are not expected to quantify parameter uncertainty accurately as pseudo-likelihoods are not generally true likelihoods \cite{ribatet_bayesian_2012, miller_asymptotic_2021}. In a complex survey sampling context, both Williams and Savitsky \cite{williams2021} and Han and Wellner \cite{han2021} consider pseudo-Bayesian inference, noting that ``vanilla" credible sets for model parameters based on a pseudo-posterior distribution do not generally converge on valid frequentist confidence intervals. The need to rescale pseudo-posterior distributions for pairwise likelihood analysis with survey data is also discussed by Thompson et al. \cite{thompson_bayesian_2022}. 

Given a correctly specified parametric superpopulation model Williams and Savitsky \cite{williams2021} observe that under certain assumptions for the sampling design and the model likelihood, the pseudo-MLE, denoted $\hat{\theta}_{\nu}^\pi$ and defined as the estimator obtained by maximizing the frequentist pseudo-likelihood, is asymptotically Gaussian. More generally, the pseudo-MLE converges to the population MLE, so in the case of model misspecification, the pseudo-MLE converges asymptotically to the Kullback-Leibler divergence minimizing parameters $\theta^*$, assuming such a vector $\theta^*$ exists in the interior of the parameter space.  In particular, $\sqrt{N_\nu}(\hat{\theta}_{\nu}^\pi-\theta^*)$ converges asymptotically to a Gaussian random variable with mean $0$ and variance $H_{\theta^*}^{-1}J_{\theta^*}^\pi H_{\theta^*}^{-1}$ where $H_{\theta^*}$ is the Fisher information:  

\begin{equation}
H_{\theta^*}=-\frac{1}{N_\nu}\sum_{j\in U_\nu}\mathbb{E}_{P_{\theta^*}}\ddot{\ell}_{\theta^*}(y_j)
\end{equation}
where $\ell_{\theta^*}=\log p_{\theta^*}$ denotes the log-likelihood, and $J_{\theta^*}^\pi$ is the variance matrix of the score functions under the combined measure $\mathbb{P}$:
\begin{equation}
J_{\theta^*}^\pi=\mathbb{E}_{\mathbb{P}}\left[\mathbb{P}_\nu^\pi\dot{\ell}_{\theta^*}\dot{\ell}_{\theta^*}^T\right]
\end{equation}
Moreover, under their set of regularity conditions,  Williams and Savitsky derive the asymptotic distribution of the pseudo-posterior:
\begin{equation}
\sup_B\bigg|\Pi_{\nu}^\pi(B\mid D^{(\nu)})-\mathcal{N}_{\hat{\theta}_{\nu}^\pi, N_\nu^{-1}H_{\theta^*}^{-1}}(B)\bigg|\rightarrow 0
\end{equation}
where $\hat{\theta}_{\nu}^\pi$ is the pseudo-MLE and $H_{\theta^*}^{-1}$ is the Fisher information.

Based on the differing forms of the covariance matrices for the pseudo-MLE and pseudo-posterior, Williams and Savitsky propose adjusting samples as follows:
\begin{equation}
\widehat{\theta}^{WS(k)}=\left(\widehat{\theta}^{(k)}-\overline{\theta}\right)R_2^{-1}R_1+\overline{\theta}
\end{equation}
where $R_1^TR_1=H_{\theta^*}^{-1}J_{\theta^*}^\pi H_{\theta^*}^{-1}$ and $R_2^TR_2=H_{\theta^*}^{-1}$. Here, $\widehat{\theta}^{(k)}$ is the $k$th sample from the pseudo-posterior and $\hat{\theta}^{WS(k)}$ the $k$th adjusted sample. Finally, $\overline{\theta}$ is the mean of the pseudo-posterior draws, but could be replaced with the pseudo-MLE. The authors call $R_2^{-1}R_1$ a multivariate design effect adjustment, which will vanish for a SRS.

In practice, $H_{\theta^*}$ is estimated as the observed information, i.e. the negative Hessian of the weighted log-likelihood at the pseudo-MLE. Following Williams and Savitsky, we estimate $J_{\theta^*}^\pi$ via a resampling approach \cite{preston_rescaled_2009} that seeks to estimate the variance of the score functions by sampling PSUs with replacement from the sample. We use numerical differentiation when estimating both $H_{\theta^*}$ and $J_{\theta^*}^\pi$. Further details are provided in the Appendix.

\subsection{Rescaling small area estimates}

Under the hierarchical model (\ref{eq-bhm}), this rescaling approach enables us to produce credible sets for the parameter vector $\theta=(\beta_0, \boldsymbol\beta_1,\sigma_u^2, \sigma_\varepsilon^2)$ that converge on asymptotically correct frequentist confidence sets for the true model parameters. However the interpretation is less clear for rescaled credible sets of area level random quantities such as $u_i$. For the purpose of small area estimation, we are interested in between area variations that are stable as $\nu\rightarrow\infty$. As such, we propose a strategy that rescales the pseudo-posterior distributions for $\beta_{0i}=\beta_0+u_i$ based on the asymptotic distributions of the pseudo-MLEs resulting from likelihood analysis of the model treating the $\beta_{0i}$ parameters as fixed effects.

In practice, for $k = 1,\ldots, K$, we draw samples $\beta_0^{(k)},\boldsymbol\beta_1^{(k)},\sigma_u^{2(k)}, \sigma_\varepsilon^{2(k)}$, and $\mathbf{u}^{(k)}$ from the pseudo-posterior distribution based on the hierarchical model (\ref{eq-bhm}). We can then transform these samples to express them in terms of the parameters of the model (\ref{eq-bhf-alt}) with fixed area-specific intercepts, yielding a sample vector $\widehat{\theta}^{(k)}=(\beta_{0i}^{(k)},\boldsymbol\beta_1^{(k)},\sigma_\varepsilon^{2(k)})$. We then estimate the rescaling matrices $H_{\theta^*}$ and $J_{\theta^*}^\pi$ using the likelihood arising from (\ref{eq-bhf-alt}), treating $\beta_{0i}$ as fixed parameters. In other words, the model we use for rescaling is a fixed effects model and the asymptotic distribution of the pseudo-MLE is based on a model that treats $\beta_{0i}$ as stable across populations. From a Bayesian standpoint, this perspective shift is natural as the distinction between fixed and random effects is less salient: we are simply defining a hierarchical prior on the parameters of interest $u_i$.

\section{Simulations}\label{s:sims}

To assess the performance of our pseudo-Bayesian approach we carry out a simulation study using a range of population models and sampling designs. For each choice of population model, we generate a single finite population of responses. Using each design, we repeatedly sample a subset of responses and then compute estimators of $\mu_i$, which we compare with the finite population means $\overline{Y}_i$

\subsection{Population generating models}

We carry out simulations for populations of continuous response data and binary response data generated using the models described below. For both response models, we first generate auxiliary variables for a clustered population letting $\mathbf{z}_{icj}$ denote the auxiliary variables for individual $j$ in cluster $c$ in area $i$. We assume that each unit belongs to one cluster and clusters are nested within areas. Finally, we assume area $i$ contains $N_C(i)$ clusters indexed by the set $C(i)=\{c_{i_1},\ldots, c_{i_{N_C(i)}}\}$.

\begin{enumerate}
	\item We generate $z_{1icj}\stackrel{ind}{\sim}  N\left(\frac{i}{m}, 1\right)$, where $1\leq i\leq m$ indexes the area, so the mean of $z_{1icj}$ varies across areas.
	\item We generate $z_{2icj}=\frac{i}{m}+z'_{2icj}$, where $z'_{2icj}\stackrel{iid}{\sim}\mathrm{Exp}(1/2)$. The variable $z_{2icj}$ represents a measure of unit size, which  we will use to specify a sampling design. We define $z_{2ic\cdot}=\sum_jz_{2icj}$ to be the cluster size obtained by summing the sizes for all units in the relevant cluster. We define the scaled unit size $x_{2ij}$ to be equal to $z_{2icj}$ scaled to have mean zero and variance $1$. Similarly, we define the scaled cluster size $\widetilde{x}_{2ij}$ to be equal to $z_{2ic\cdot}$ scaled to have mean zero and variance $1$, where $c$ is the cluster containing unit $j$.
\end{enumerate}

\subsubsection{Continuous responses}

To generate continuous response data, we simulate data from population models of the form 
\begin{equation}
y_{ij}=\beta_0+\beta_1x_{1ij}+\beta_2\widetilde{x}_{2ij}+\varepsilon_{ij}
\end{equation}
where $\varepsilon_{ij}\stackrel{iid}{\sim} N(0,\sigma_\varepsilon^2)$. As described above, $\widetilde{x}_{2ij}$ denotes the cluster size, representing a relevant design variable that is unavailable to the analyst. To estimate the area level means, we fit the following nested error regression model:
\begin{equation}
y_{ij}=\beta_0'+\beta_1'x_{1ij}+u_i+\varepsilon_{ij}'=\beta_{0i}'+\beta_1'x_{1ij}+\varepsilon_{ij}'
\end{equation}
For the purpose of estimating model parameters, we assume that $\varepsilon_{ij}'\stackrel{iid}{\sim} N(0,\sigma_\varepsilon^2)$. Note that for the estimation model, we use $\beta_0', \beta_1',$ and $\varepsilon_{ij}'$ to denote model parameters to emphasize that this model is misspecified and we cannot expect to obtain consistent estimators for the true population parameters in general. Based on this estimation model, $u_i$ is used to capture stable between area differences induced by area level differences in the mean value of $\widetilde{x}_{2ij}$.  
\subsubsection{Binary responses}

We also implement simulations for binary response data using the population generating models of the form:
\begin{equation}
\begin{split}
y_{ij}\mid q_{ij}&\sim \mathrm{Bernoulli}(q_{ij})\\
q_{ij}&=\mathrm{expit}(\beta_0+\beta_1x_{1ij}+\beta_2\widetilde{x}_{2ij})
\end{split}
\end{equation}
We again use  $\widetilde{x}_{2ij}$ to denote the unobserved cluster size for individual $j$. The pseudo-Bayesian approach described above can be adapted to non-Gaussian response data by using alternative likelihoods. To estimate the area level means, we fit the following logistic regression model:
\begin{equation}
\begin{split}
y_{ij}\mid q_{ij}&\sim \mathrm{Bernoulli}(q_{ij})\\
q_{ij}\mid u_i&=\mathrm{expit}(\beta_0'+\beta_1'x_{1ij}+u_i)=\mathrm{expit}(\beta_{0i}'+\beta_1'x_{1ij})\\
u_i&\stackrel{iid}{\sim} N(0,\sigma_u^2)\label{e:est-log-reg}
\end{split}
\end{equation}
The areal effects $u_i$ capture between area differences in the log-transformed odds induced by area level differences in the mean value of $\widetilde{x}_{2ij}$. Again, we can either view $\beta_{0i}'=\beta_{0}'+u_i$ as a fixed area specific intercept term or as a random intercept by placing a Gaussian prior on $u_i$. Based on this estimation model, the target of estimation is defined as 
\begin{equation}
\mu_i=\mathbb{E}_{\mathbb{P}}\left[\mathrm{expit}(\beta_{0i}'+\beta_1'x_{1ij})\right]
\end{equation}
where the expectation is taken with respect to both the model and design. 

\subsection{Estimation procedure}

Based on our estimation models, we consider three approaches for estimating area level means. First, we consider a Bayesian approach ignoring the weights (\textbf{Unwt}) and treating sampling as ignorable. Next, we implement our pseudo-Bayesian approach using the sampling weights, both with (\textbf{WtRscl}) and without (\textbf{Wt}) the rescaling step described in the previous section. The sampling weights used in this analysis are normalized so that their sum is equal to the observed sample size $n$. For all of these Bayesian estimators, we compute posterior medians and 90\% credible sets. We compare these three Bayesian estimators with two design-based estimators: the \textbf{H\'ajek} estimator and a generalized regression estimator (\textbf{GREG}) based on a working model with fixed area specific intercepts. For continuous data, this working model takes the form $y_{ij}=\beta_{0i}'+\beta_1'x_{1ij}+\varepsilon_{ij}'$. We compute 90\% prediction intervals based on the estimated mean squared predictive error of these estimators.

We approximate the unscaled pseudo-posterior distributions using integrated nested Laplace approximation, as implemented in the INLA package \cite{rue2017}, which facilitates fast approximate Bayesian inference for latent Gaussian models as an alternative to other methods such as Markov chain Monte Carlo. Using INLA, we can obtain samples from the pseudo-posterior distributions for the parameters of interest. Further detail on the estimation procedures, including descriptions of priors for model hyperparameters can be found in the Appendix. Code used to produce the results throughout this manuscript can be found on GitHub.

Since $\widetilde{x}_{2ij}$ is unobserved, our estimation models are misspecified. As noted by Williams and Savitsky \cite{williams2021}, their asymptotic results rely on correct parameterization of the dependence structure for the population. However, the unobserved $\widetilde{x}_{2ij}$ is constant within clusters, and induces cluster dependence in our observations. Simulations by Williams and Savitsky indicate that their proposed rescaling method may be robust to this misspecification. We consider an asymptotic framework in which the number of clusters sampled in each area, but not the size of the clusters, is increasing, but we are primarily interested in the performance of these estimators in a small sample setting, which will be of practical relevance to small area estimation.

\subsection{Sampling designs}

We consider different sampling designs which induce dependence between observed response values, some of which are informative with respect to the analyst-specified model.  For all designs, we stratify sampling by the $m$ small areas.
\begin{enumerate}
	\item \textbf{Stratified random sampling without replacement (SRS)} Within each area $i$, we sample $n(i)$ individuals at random without replacement. Under this design, assuming the sampling fraction is small, the design effect is expected to be small, making the effect of incorporating sampling weights during estimation negligible.
	\item\textbf{Single stage informative sampling (PPS1)} For this design, within each area, we sample $n(i)$ units without replacement, with probability proportional to size $s_{ij}=x_{2ij}-\min(x_{2ij})+1$ using Midzuno's method as implemented in the R package \texttt{sampling} \cite{tille2021}. This yields an single stage design with unequal sampling probabilities (PPS1) that is informative with respect to the analyst-specified model since $s_{ij}$ is correlated with the unobserved cluster size $\widetilde{x}_{2ij}$.
	\item\textbf{Two stage informative sampling (PPS2)} Within each area $i$, we then sample $n_C(i)$ clusters without replacement with probability proportional to size $\widetilde{x}_{2ij}-\min(\widetilde{x}_{2ij})+1$. Within each sampled cluster, we sample $n(i,c)$ units with probability proportional to size $s_{ij}=x_{2ij}-\min(x_{2ij})+1$. This yields a two stage design with unequal sampling probabilities (PPS2) that is informative with respect to the model since $\widetilde{x}_{2ij}$ is unobserved.
\end{enumerate}
\subsection{Results}

For each data generating model, we generate auxiliary variables for a finite population consisting of $N=90,000$ individuals divided evenly between $m=20$ areas. Each area is divided into $N_C=150$ clusters of thirty individuals. Based on this fixed auxiliary data, we repeatedly simulate response data and sample from the resulting population for each sampling design for a total of 1,000 simulations. For the continuous response case, we simulate data from the following model:
\begin{equation}
y_{ij}=x_{1ij}+2\widetilde{x}_{2ij}+\varepsilon_{ij}
\end{equation}
where $\varepsilon_{ij}\stackrel{iid}{\sim} N(0,1)$. 
For the binary response case, we simulate population data from the following model:
\begin{equation}
\begin{split}
y_{ij}\mid q_{ij}&\sim \mathrm{Bernoulli}(q_{ij})\\
q_{ij}&=\mathrm{expit}(x_{1ij}+2\widetilde{x}_{2ij})
\end{split}
\end{equation}
We compute the finite population area level mean $\overline{Y}_i$ for each area $i$. In each simulation, we compute point estimates $\widehat{\mu}_i$ as well as 90\% interval estimates $(\widehat{\mu}_i^-,\widehat{\mu}_i^+)$ for every estimator. For each method, we compute root mean squared error (RMSE) and mean absolute error (MAE). We also compute the emprical coverage of the 90\% interval estimates and the mean interval lengths (MIL) across all areas, averaged across all simulations.
\begin{align}
\mathrm{RMSE}(\widehat{\boldsymbol\mu})&=\sqrt{\frac{1}{m}\sum_{i}(\overline{Y}_i-\widehat{\mu}_i)^2}\\
\mathrm{MAE}(\widehat{\boldsymbol\mu})&=\frac{1}{m}\sum_{i}|\overline{Y}_i-\widehat{\mu}_i|\\
\mathrm{Cov}_{90}(\widehat{\boldsymbol\mu})&=\frac{1}{m}\sum_{i}\mathbf{1}\{\overline{Y}_i\in(\widehat{\mu}_i^-,\widehat{\mu}_i^+)\}\\
\mathrm{MIL}_{90}(\widehat{\boldsymbol\mu})&=\frac{1}{m}\sum_{i}(\widehat{\mu}_i^+ - \widehat{\mu}_i^-)
\end{align}

\input{figures/cts_simulation_summary.tex}

\input{figures/cts_simulation_summary_large_n.tex}

Table \ref{tab:cts-sim-smry} summarizes the continuous response simulation results for a small sample setting where $n(i)=30$ and Table \ref{tab:cts-sim-smry-lg} provides an analogous summary for a larger sample setting where $n(i)=100$. For the PPS2 design, we assume that $n(i)/5$ clusters are sampled and five individuals are sampled within each cluster. 

Under stratified random sampling, the model-based (Unwt, Wt, and WtRscl) approaches perform similarly and produce point estimates with the lowest RMSE and MAE, and achieve close to nominal interval coverage rates. For the large sample simulations, the model-based estimates perform similarly to the GREG estimator, indicating the reduced shrinkage compared with the small sample case.

Under the PPS1 sampling, in the small sample simulations, the weighted model-based methods perform best in terms of point estimates, with the weighted and rescaled estimates (WtRscl) yielding slightly wider interval estimates on average. For the large sample simulations, the unweighted and weighted but not rescaled interval estimates exhibit undercoverage while the weighted and rescaled intervals are better calibrated.

Finally, for the PPS2 sampling design, the weighted and rescaled method achieves the best performance in terms of RMSE and MAE as well as calibrated interval coverage. Under this design, the other model-based methods exhibit large undercoverage.

\input{figures/binary_simulation_summary.tex}

\input{figures/binary_simulation_summary_large_n.tex}

Table \ref{tab:bin-sim-smry} summarizes simulation results for binary response data with sample size $n(i)=30$ and Table \ref{tab:bin-sim-smry-lg} provides an analogous summary with sample size $n(i)=100$. The results from these simulations are similar to those from the continuous case under the SRS and PPS2 designs, with the weighted and rescaled estimates generally producing the best point estimates and interval estimates with close to nominal coverage. Under the PPS1 design, the benefits of rescaling are less clear, but the weighted and rescaled method does not perform significantly worse than the other model-based approaches.

\section{Application}\label{s:applications}

We apply the pseudo-Bayesian approach to estimate measles vaccination coverage for prefectures in Guinea in 2018 based on data from the Demographic and Health Surveys (DHS) Program. The DHS Program conducts surveys in many LMIC, typically using a stratified two-stage cluster sampling design. Each country is first divided by its principal administrative regions, usually called Admin-1 regions. Each region is partitioned into urban and rural components. Sampling is stratified by crossing the Admin-1 regions with urban/rural labels. For the first stage of sampling, each stratum is divided into clusters, or enumeration areas (EAs). Within each stratum, a pre-specified number of clusters is sampled with probability proportional to size. In the second stage, a fixed number of households is sampled from each selected cluster. Under this sampling design, cluster size is a relevant design variable that is typically not made public but which could be associated with the response.

We estimate subnational vaccination rates for the first dose of measles-containing-vaccine (MCV1) among children aged 12-23 months in Guinea using data from the 2018 Guinea DHS \cite{institut_national_de_la_statistique_and_icf_guinea_2019}. This survey interviewed mothers in each selected household and collected vaccination data for their children based on vaccination cards or caregiver recall. We produce estimates for each prefecture, which correspond to subdivisions of Guinea's eight Admin-1 regions.  We refer to these prefectures as Admin-2 regions. We rely on the boundaries published by Database of Global Administrative Areas (GADM) \cite{noauthor_database_2022}.

\begin{figure}
	\centering
	\includegraphics[scale=.45]{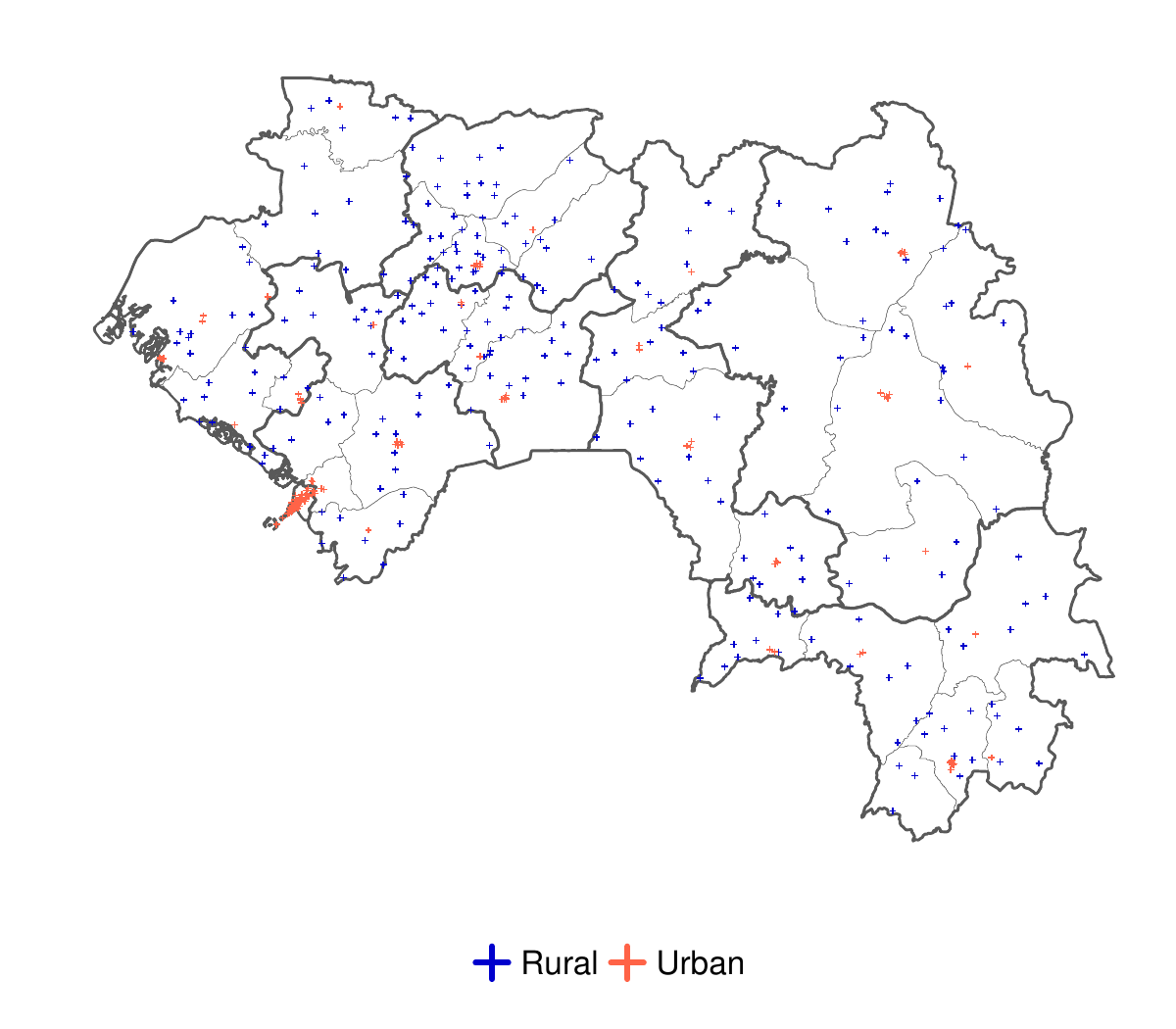}
	\caption{Map of Guinea with Admin-1 level boundaries (thick borders) and Admin-2 level boundaries (thin borders). Points indicate enumeration area locations for which data on measles vaccination is available.}
	\label{fig:map}
\end{figure}

The design for the 2018 DHS was based on a sampling frame created using data from a 2017 census which identified 9679 enumeration areas divided into 15 strata (from splitting eight Admin-1 areas into urban/rural components minus the entirely urban zone of Conakry). Data were collected from 401 clusters. The DHS Program publishes coordinates for all selected clusters after displacing their locations by small distances to protect privacy. Figure \ref{fig:map} provides the Admin-1 and Admin-2 boundaries and displaced EA locations in Guinea for which data were collected.

We generate estimates using design-based methods and unit level logistic regression models. For the unit level models, we use two covariates based on estimated travel times to cities in 2015 \cite{weiss_global_2018} and the intensity of night time lights as observed via satellite imagery in 2016 \cite{worldpop_global_2018}. Note that these covariates are themselves estimated using statistical modeling. We also use estimated population counts produced by WorldPop \cite{worldpop_global_2020} to create a binary covariate classifying pixels as either urban or rural by the highest population pixels in an area to be urban so that the proportion of individuals classified urban equals that which is reported in the 2018 Guinea DHS report.

We use a logistic regression model, so covariate information for the entire population is required to generate estimates for each individual. Since complete population data is not available, instead of making a separate prediction for each child, we make predictions for each pixel and aggregate these predictions to get an area level estimate of the mean outcome of interest. When aggregating, we weight each pixel by its estimated age 1-5 population \cite{worldpop_age_2018}. We project all covariate values to the 1km by 1km grid used by WorldPop. 

\begin{figure}
	\centering
	\includegraphics[scale=.45]{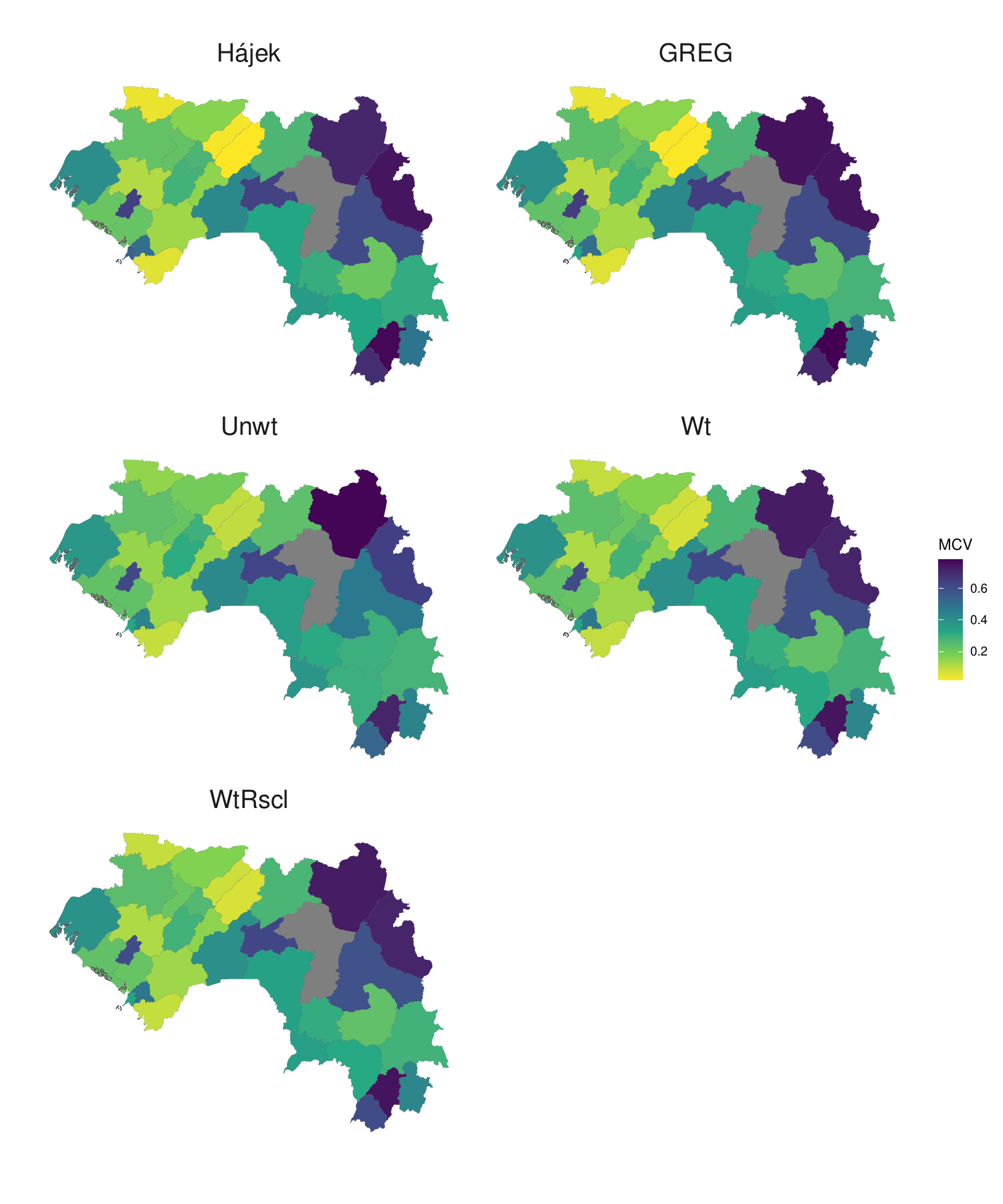}
	\caption{Estimated measles vaccination rates among children aged 12-23 months for Admin-2 areas in Guinea in 2018.}
	\label{fig:mcv-ests}
\end{figure}

\begin{figure}
	\centering
	\includegraphics[scale=.499]{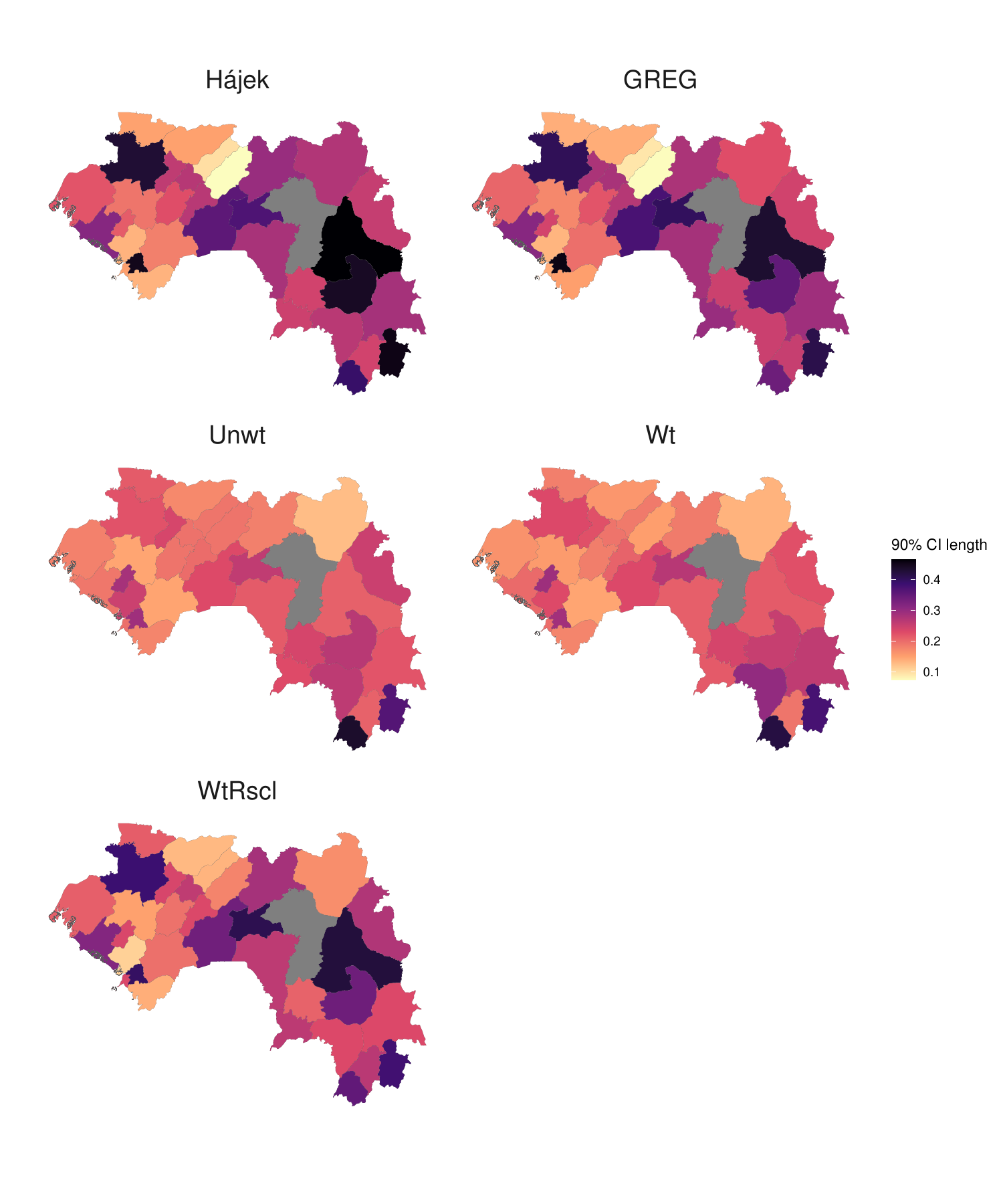}
	\caption{Prediction interval lengths for estimated measles vaccination rates for Admin-2 areas in Guinea in 2018.}
	\label{fig:mcv-lens}
\end{figure}

Figure \ref{fig:mcv-ests} compares point estimates of measles vaccination rates and \ref{fig:mcv-lens} provides the length of interval estimates among children aged 12-23 months for Guinea's prefectures in 2018. We provide point and interval estimates for all methods in the Appendix. In general, the point estimates produced by all methods are quite similar, but the estimates of uncertainty vary considerably. In general, the unweighted Bayes and weighted but not rescaled Bayes estimates have the shortest prediction intervals, indicating the least uncertainty. The design-based H\'ajek and GREG approaches produce longer prediction intervals. The weighted and rescaled method generally produces larger estimates of uncertainty, generating intervals whose lengths more closely resemble those of the design-based approaches. Note that the direct H\'ajek estimate of the vaccination rate for the Admin-2 prefecture of Kouroussa is zero and thus a direct estimator of the associated variance is unavailable. As a result, we omit estimates for Kouroussa, which is depicted in gray in Figures \ref{fig:mcv-ests} and \ref{fig:mcv-lens}.

\section{Discussion}\label{s:dis}

Pseudo-Bayesian approaches inference enables analysts to leverage available sampling weights to adjust for features of the survey design that cannot be incorporated into a small area estimation model. However, since credible sets based on a naive pseudo-posterior exhibit undercoverage \cite{leon-novelo_fully_2019}, the pseudo-posterior must be rescaled to produce credible sets that quantify uncertainty meaningfully. Using a rescaling post-processing adjustment proposed by Williams and Savitsky \cite{williams2021}, we show that pseudo-Bayesian approaches can be used to generate improved point and interval estimators for small area means of continuous and binary outcomes.

Previous applications of pseudo-Bayesian approaches rely on scaling the sampling weights to sum to the sample size as an ad hoc solution for scaling the pseudo-posterior. The approach proposed by Williams and Savitsky first scales the sampling weights but then also estimates a multivariate design effect for the parameters of interest using the available data, which is subsequently used to rescale the pseudo-posterior. Both the initial scaling of sampling weights and the rescaling of parameter samples are potentially valuable. The initial scaling of sampling weights controls degree of shrinkage induced by the Gaussian prior on the random effects. If the unscaled weights are used, then the degree of shrinkage may be too low because inference on the random effects proceeds as though a population of size $N$ is observed. The subsequent rescaling of the samples from the pseudo-posterior is aimed at improving the coverage of credible sets for parameters of interest. Han and Wellner \cite{han2021} propose a similar rescaling approach but do not explicitly encourage rescaling the sampling weights when defining the pseudo-posterior.

A key limitation of the approach presented here is that we focus on estimation targets that can be expressed in terms of fixed parameters for which we have asymptotically increasing amounts of data. Our approach relies upon having sufficient data to rescale model-based estimates of uncertainty, so as observed in Section \ref{s:applications}, when we have severely limited or no data in an area, we are unable to construct valid prediction intervals. When using unit level models for small area estimation, it has become increasingly common to model outcomes of interest as continuous spatial processes and generate area level estimates by aggregating predictions made on a high-resolution spatial grid. Under such an approach, predictions may be required for each individual cluster or location. Our approach does not account for the case in which the targets of estimation are themselves observed at random. For example, we do not seek to estimate individual cluster level effects. Aggregating unit level model predictions for individual clusters or pixels to obtain area level predictions is difficult and requires careful consideration of the sampling design \cite{paige2022, paige_spatial_2022}.

\bibliographystyle{ieeetr}
\bibliography{svyulm.bib}

\appendix 
\section{Conditions for asymptotic normality of the pseudo-posterior}

We rely upon a result from Kleijn and Van der Vaart \cite{kleijn_bernstein-von-mises_2012} that establishes conditions for asymptotic normality of a posterior distribution under model misspecification. This result has previously been applied in the context of complex sampling by Han and Wellner \cite{han2021} and Williams and Savitsky \cite{williams2021}. However, both Han and Wellner and Williams and Savitsky only consider model misspecification caused by the complex sampling procedure, assuming that the analyst-specified model correctly describes the superpopulation. In other words, if the model can be indexed by parameter vector $\theta\in \Theta$, the true superpopulation distribution is given by $P_{\theta_0}$. In our application, we assume that the true population generating distribution $P_0$ may not belong to the specified model. Under this form of misspecification, Kleijn and Van der Vaart show that the posterior distribution concentrate asymptotically on a parameter vector that minimizes the Kullback-Leibler distance between the true distribution and the model, which we denote $\theta^*\in \Theta$.

We rely on Han and Wellner's extension of a result from Kleijn and Van der Vaart \cite{kleijn_bernstein-von-mises_2012} for pseudo-posterior distributions. We update the conditions of Han and Wellner \cite{han2021} to reflect potential model misspecification of the finite population model. 

\begin{prop}
	Assume the following conditions hold for some parameter vector $\theta^*\in\Theta$.
	\begin{enumerate}
		\item (Stochastic local asymptotic normality (LAN) condition) There exist random vectors $\Delta_{\nu,\theta^*}$ and a non-singular matrix $H_{\theta^*}$ such that $\Delta_{\nu,\theta^*}$ is bounded in probability and for every compact $K\subset \mathbb{R}^d$,
		$$\sup\limits_{h\in K}\Bigg|N_\nu \mathbb{P}^\pi_{N_\nu}\log\frac{p_{\theta^*+ h/\sqrt{N_\nu}}}{p_{\theta^*}}-h^TH_{\theta^*}\Delta_{\nu,\theta^*}-\frac{1}{2}h^TH_{\theta^*}h\Bigg|=o_{\mathbb{P}}(1)$$
		\item (Sufficient prior mass condition) The prior $\Pi$ on $\Theta$ has Lebesgue density $\pi$ that is continuous and positive on a neighborhood of $\theta^*$.
		\item (Posterior contraction rate). For every sequence of constants $L_\nu\rightarrow \infty$, 
		$$\mathbb{P} \Pi_{N_\nu}^\pi\left(\theta\in \Theta:||\theta-\theta^*||>L_{\nu}/\sqrt{N_\nu}\mid  D^{(\nu)}\right)\rightarrow 0$$
	\end{enumerate}
	Then, the sequence of pseudo-posterior distributions $ \Pi_{N_\nu}^\pi$ converges in total variation to a sequence of normal distributions:
	
	$$\sup_B\Bigg| \Pi_{N_\nu}^\pi(\sqrt{N_\nu}(\theta-\theta^*)\in B\mid D^{(\nu)})-\mathcal{N}_{\Delta_{\nu,\theta^*}, H_{\theta^*}^{-1}}(B)\Bigg|=o_{\mathbb{P}}(1)$$
\end{prop}
\begin{lem} Assume the following conditions:
	\begin{enumerate} 
		\item  For some nonrandom $\pi_0>0$, $$\min_{1\leq j\leq N_\nu}\pi_j\geq \pi_0$$
		\item The weights satisfy a central limit theorem $$\frac{1}{\sqrt{N_\nu}}\sum_{j=1}^N\left(\dfrac{\delta_j}{\pi_j}-1\right)=\mathcal{O}_{\mathbb{P}}(1)$$
		\item The map $\theta\mapsto\log p_\theta(x)=\ell_\theta(x)$ is differentiable at $\theta^*$ for all $x$ with derivative $\dot\ell_{\theta^*}(x)$, and for $\theta_1,\theta_2$ close enough to $\theta^*$
		\[|\ell_{\theta_1}(x)-\ell_{\theta_2}(x)|\leq m(x)||\theta_1-\theta_2||\]
		for some $P_0$ square integrable function $m$.
		\item The Kullback-Leibler divergence relative to the true superpopulation measure $P_0$ has a second-order Taylor-expansion:
		\[
		P_0\log\dfrac{p_{\theta^*}}{p_\theta}=\frac{1}{2}(\theta-\theta^*)^\top H_{\theta^*}(\theta-\theta^*)+o(||\theta-\theta^*||^2)
		\]
		where $H_{\theta^*}$ is a positive-definite Hessian matrix.
	\end{enumerate}
	Then the stochastic LAN condition holds where $\Delta_{\nu,\theta^*}=H_{\theta^*}^{-1}\mathbb{G}_{N_\nu}^\pi \dot{\ell}_{\theta^*}$.
\end{lem}

It remains to show the pseudo-posterior contracts at a $\sqrt{N_\nu}$ rate. Han and Wellner provide conditions under which the pseudo-posterior contracts at the desired rate assuming the population generating distribution belongs to the analyst-specified model. Kleijn and van der Vaart provide more general conditions under which a posterior distribution contracts at the desired rate, under misspecification, which may be combined with results from Savitsky and Toth \cite{savitsky_bayesian_2016} ensuring the pseudo-posterior distribution converges to the posterior distribution based on the full finite population.

\section{Misspecification of the superpopulation model}

For our estimation models, the parameter vector $\theta$ is composed of the coefficients $\boldsymbol\beta_0$, $\boldsymbol\beta_1$ and potentially a residual variance parameter. However, we are primarily interested in pseudo-posterior samples of the small area means $\mu_i$, $i=1,\ldots, m$, which we obtain by transforming the pseudo-posterior samples of $\theta$. Under misspecification, the pseudo-posterior for $\theta$ concentrates asymptotically on the Kullback-Leibler divergence minimizing parameter vector $\theta^*$, which may not generally be of interest. In this section, we show that in simple cases, even when the estimation model is misspecified, pseudo-posterior samples of $\theta$ can be used to obtain reasonable estimates of the finite population small area means.

\subsection{Linear regression}

Under the fixed intercepts linear nested error regression model, we can derive closed-form expressions for the census maximum likelihood estimators:
\[
\widehat{\boldsymbol\beta_1}=(\mathbf{X}^T\mathbf{X})^{-1}\mathbf{X}^T\mathbf{Y}
\]
\[
\widehat{\beta}_{0i}=\overline{Y}_i-\overline{\mathbf{X}}_i^T\widehat{\boldsymbol\beta_1}
\]
where $\mathbf{X}$ denotes the population covariate matrix and $\mathbf{Y}$ denotes the population response vector. As a result, we can define a model-based predictor of $\mu_i$ as follows:
\[\widehat{\mu_i}=\widehat{\beta}_{0i}+\overline{\mathbf{X}}_i^T\widehat{\boldsymbol\beta_1}=\overline{Y}_i,\]
so the population estimate of $\mu_i$ coincides with the true finite population mean $\overline{Y}_i$.
In this sense, assuming the pseudo-posterior concentrates on the census maximum likelihood estimators, even if the model is misspecified, the pseudo-posterior for a small area mean will concentrate on the true  finite population mean because the model includes area-specific fixed intercept terms. For example, if there are missing relevant covariates, the estimates of $\boldsymbol\beta_1$ may be biased, but the resulting small area mean estimates can be robust to misspecification.

\subsection{Logistic regression}

Closed-form expressions will not generally be available for the maximum likelihood estimators for nested error regression models with non-Gaussian likelihoods. Below, we consider how misspecification of a logistic regression model of the form in (\ref{e:est-log-reg}) affects the resulting small area estimators. Given parameter estimates for $\beta_{0i}$ and $\boldsymbol\beta_1$, a model-based predictor of $\mu_i$ can be computed if covariate information is available for the full population:
\[\widehat{\mu}_i =\frac{1}{N(i)}\sum_{j\in U(i)}\mathrm{expit}(\widehat{\beta}_{0i}+\mathbf{x}_{ij}^T\widehat{\boldsymbol\beta_1})=\frac{1}{N(i)}\sum_{j\in U(i)}\widehat{q}_{ij}\]
Given full population data, an estimating equations approach for estimating $\beta_{0i}$ and $\boldsymbol\beta_1$ involves solving the following sets of equations for $q_{ij}=\mathrm{expit}(\beta_{0i}+\beta_1x_{1ij})$:
\begin{align}
	\sum_{j\in U(i)}(y_{ij}-q_{ij})&=0;\; i=1,\ldots, m\label{e:logit-reg-cdn}\\
	\sum_{i}\sum_{j\in U(i)}(y_{ij}-q_{ij})\mathbf{x}_{ij}&=0
\end{align}

As such, an approximate solution to these equations will yield an estimator $\widehat{\mu}_i$ that is close to $\overline{Y}_i$ due to (\ref{e:logit-reg-cdn}), which requires that the sum of the predicted $q_{ij}$ values for units in area $i$ sums to the true finite population total.

\section{Estimating the multivariate design effect}

As described above, Williams and Savitsky propose a post-processing adjustment that scales the pseudo-posterior by rescaling samples:
\begin{equation}
\widehat{\theta}^{WS(k)}=\left(\widehat{\theta}^{(k)}-\overline{\theta}\right)R_2^{-1}R_1+\overline{\theta}
\end{equation}
where $\widehat{\theta}^{(k)}$ is the $k$th sample from the pseudo-posterior, $\hat{\theta}^{WS(k)}$ the $k$th adjusted sample,  $\overline{\theta}$ is the vector mean of the samples $\widehat{\theta}^{(k)}$, $R_1^TR_1=H_{\theta^*}^{-1}J_{\theta^*}^\pi H_{\theta^*}^{-1}$, and $R_2^TR_2=H_{\theta^*}^{-1}$. We define $H_{\theta^*}$ as
\begin{equation}
H_{\theta^*}=-\frac{1}{N_\nu}\sum_{j\in U_\nu}\mathbb{E}_{P_{\theta^*}}\ddot{\ell}_{\theta^*}(y_j, \mathbf{x}_j)
\end{equation}
and $J_{\theta^*}^\pi$ is the variance matrix of the weighted score functions under $\mathbb{P}$:
\begin{equation}
J_{\theta^*}^\pi=\mathbb{E}_{{\theta^*}, \nu}\left[\mathbb{P}_\nu^\pi\dot{\ell}_{\theta^*}\dot{\ell}_{\theta^*}^T\right]
\end{equation}
 In practice, the matrices $H_{\theta^*}$ and $J_{\theta^*}^\pi$ can be estimated via Algorithm 1 from Williams and Savitsky \cite{williams2021}. Based on the model (\ref{eq-bhf-alt}) with fixed area-specific intercepts, we obtain sample vectors $\widehat{\theta}^{(k)}=(\beta_{0i}^{(k)},\boldsymbol\beta_1^{(k)},\sigma_\varepsilon^{2(k)})$. We can then compute the mean of the pseudo-posterior draws $\overline{\theta}$. We compute $\widehat{H}_{\theta^*}$ as the negative Hessian of the weighted log-likelihood for model (\ref{eq-bhf-alt}) with fixed $\beta_{0i}$ at the pseudo-MLE.
 
We estimate $J_{\theta^*}^\pi$ using a resampling approach that repeatedly subsamples PSUs without replacement, within strata, from the sample \cite{preston_rescaled_2009}. For each subsample, we can compute a weighted score function. We then compute the sample covariance of the weighted score functions, across 100 subsamples, to obtain an estimator $\widehat{J}_{\theta^*}^\pi$. Finally, we can compute $\widehat{R}_1$ and $\widehat{R}_2$ via Cholesky decomposition, plugging in $\widehat{H}_{\theta^*}$ and $\widehat{J}_{\theta^*}^\pi$ in the definitions of $R_1^TR_1$ and $R_2^TR_2$.

\section{Estimation procedure}

The simulations and application described above were implemented in the R programming language \cite{r_core_team_r_2023}. For general data cleaning and processing covariate information, we used the \texttt{tidyverse} \cite{tidyverse}, \texttt{sf} \cite{sf}, and \texttt{terra} \cite{terra} packages. For computing the design-based H\'ajek and GREG estimators, in addition to carrying out the resampling for estimating the multivariate design, we use the \texttt{survey} package \cite{lumley2011}.

We approximate the unscaled pseudo-posterior distributions using the \texttt{INLA} package \cite{rue2017}, which enables the user to input weights when carrying out Bayesian inference. We obtain samples from the approximated pseudo-posterior distributions, and then transform and rescale the samples as described above. Credible sets are constructed by taking relevant quantiles of the rescaled samples for $\mu_i$ for all $I$.

For all unit level models, we place a flat prior on the intercept and fixed effects, so $\pi(\boldsymbol\beta_1)\propto 1$. We use penalized complexity priors for the variance parameters which place a prior on the Kullback-Leibler distance between a full model to a simplified base model, shrinking variance components $\sigma_\varepsilon$ and $\sigma_u$ to zero \cite{simpson_penalising_2017}. In particular, we specify the prior for $\sigma_u^2$ and $\sigma_\varepsilon^2$ such that $P(\sigma_u>3)=0.05$ and $P(\sigma_\varepsilon>3)=0.05$. 

Code for the simulations and the application, as well as a package with functions for fitting the models described and rescaling pseudo-posterior samples can be found on GitHub.

\end{document}

%% file: figures/cts_simulation_summary.tex
\begin{table}

\caption{\label{tab:cts-sim-smry}Averaged evaluation metrics of estimators of area level means across 1,000 continuous response simulations for SRS, PPS1, and PPS2 designs for a sample size of thirty units per area.}
\centering
\fontsize{7}{9}\selectfont
\begin{tabular}[t]{llrrrr}
\toprule
Design & Method & RMSE (x 100) & MAE (x 100) & MIL (x 100) & 90\% Int. Cov.\\
\midrule
SRS & Hájek & 38.2 & 30.4 & 125.1 & 89\\
\cmidrule{2-6}
 & GREG & 33.5 & 26.7 & 109.8 & 89\\
\cmidrule{2-6}
 & Unwt & 32.6 & 26.0 & 107.7 & 90\\
\cmidrule{2-6}
 & Wt & 32.6 & 26.0 & 107.7 & 90\\
\cmidrule{2-6}
& WtRscl & 32.6 & 26.0 & 104.7 & 88\\
\cmidrule{1-6}
PPS1 & Hájek & 41.4 & 33.0 & 133.9 & 88\\
\cmidrule{2-6}
 & GREG & 36.7 & 29.2 & 117.4 & 88\\
\cmidrule{2-6}
 & Unwt & 35.9 & 28.7 & 109.1 & 87\\
\cmidrule{2-6}
 & Wt & 35.6 & 28.4 & 107.5 & 87\\
\cmidrule{2-6}
& WtRscl & 35.6 & 28.4 & 112.0 & 87\\
\cmidrule{1-6}
 PPS2 & Hájek & 70.6 & 56.1 & 217.4 & 84\\
\cmidrule{2-6}
 & GREG & 67.6 & 53.7 & 208.1 & 84\\
\cmidrule{2-6}
 & Unwt & 72.1 & 57.3 & 105.1 & 54\\
\cmidrule{2-6}
 & Wt & 65.0 & 51.6 & 103.0 & 58\\
\cmidrule{2-6}
 & WtRscl & 65.0 & 51.6 & 200.4 & 84\\
\bottomrule
\end{tabular}
\end{table}

%% file: figures/cts_simulation_summary_large_n.tex
\begin{table}

\caption{\label{tab:cts-sim-smry-lg}Averaged evaluation metrics of estimators of area level means across 1,000 continuous response simulations for SRS, PPS1, and PPS2 designs for a sample size of one hundred units per area.}
\centering
\fontsize{7}{9}\selectfont
\begin{tabular}[t]{llrrrr}
\toprule
Design & Method & RMSE (x 100) & MAE (x 100) & MIL (x 100) & 90\% Int. Cov.\\
\midrule
SRS & Hájek & 20.8 & 16.7 & 69.0 & 90\\
\cmidrule{2-6}
 & GREG & 18.3 & 14.6 & 60.6 & 90\\
\cmidrule{2-6}
 & Unwt & 18.1 & 14.5 & 60.1 & 90\\
\cmidrule{2-6}
 & Wt & 18.1 & 14.5 & 60.1 & 90\\
\cmidrule{2-6}
 & WtRscl & 18.1 & 14.5 & 59.5 & 90\\
\cmidrule{1-6}
PPS1 & Hájek & 22.7 & 18.1 & 74.6 & 90\\
\cmidrule{2-6}
 & GREG & 20.0 & 15.9 & 65.4 & 90\\
\cmidrule{2-6}
 & Unwt & 23.2 & 18.8 & 60.9 & 81\\
\cmidrule{2-6}
 & Wt & 19.8 & 15.8 & 60.1 & 87\\
\cmidrule{2-6}
& WtRscl & 19.8 & 15.8 & 64.3 & 89\\
\cmidrule{1-6}
PPS2 & Hájek & 37.4 & 29.4 & 126.1 & 90\\
\cmidrule{2-6}
 & GREG & 35.4 & 28.1 & 120.7 & 90\\
\cmidrule{2-6}
 & Unwt & 47.5 & 39.0 & 60.5 & 44\\
\cmidrule{2-6}
 & Wt & 34.9 & 27.7 & 59.4 & 61\\
\cmidrule{2-6}
& WtRscl & 34.9 & 27.7 & 118.7 & 89\\
\bottomrule
\end{tabular}
\end{table}

%% file: figures/binary_simulation_summary.tex
\begin{table}

\caption{\label{tab:bin-sim-smry}Averaged evaluation metrics of estimators of area level means across 1,000 binary response simulations for SRS, PPS1, and PPS2 designs for a sample size of thirty units per area.}
\centering
\fontsize{7}{9}\selectfont
\begin{tabular}[t]{llrrrr}
\toprule
Design & Method & RMSE (x 100) & MAE (x 100) & MIL (x 100) & 90\% Int. Cov.\\
\midrule
SRS & Hájek & 8.1 & 6.4 & 26.4 & 89\\
\cmidrule{2-6}
 & GREG & 7.7 & 6.1 & 25.1 & 88\\
\cmidrule{2-6}
 & Unwt & 7.2 & 5.8 & 23.1 & 89\\
\cmidrule{2-6}
 & Wt & 7.2 & 5.8 & 23.1 & 89\\
\cmidrule{2-6}
& WtRscl & 7.2 & 5.8 & 23.2 & 89\\
\cmidrule{1-6}
PPS1 & Hájek & 8.9 & 7.0 & 28.5 & 87\\
\cmidrule{2-6}
 & GREG & 8.5 & 6.7 & 27.0 & 87\\
\cmidrule{2-6}
 & Unwt & 7.6 & 6.0 & 23.0 & 88\\
\cmidrule{2-6}
 & Wt & 7.8 & 6.2 & 23.1 & 87\\
\cmidrule{2-6}
& WtRscl & 7.8 & 6.2 & 25.1 & 89\\
\cmidrule{1-6}
PPS2 & Hájek & 11.8 & 9.4 & 35.5 & 80\\
\cmidrule{2-6}
 & GREG & 11.5 & 9.2 & 34.3 & 80\\
\cmidrule{2-6}
 & Unwt & 10.8 & 8.4 & 22.6 & 74\\
\cmidrule{2-6}
 & Wt & 10.3 & 8.1 & 22.9 & 75\\
\cmidrule{2-6}
& WtRscl & 10.3 & 8.1 & 31.6 & 85\\
\bottomrule
\end{tabular}
\end{table}

%% file: figures/binary_simulation_summary_large_n.tex
\begin{table}

\caption{\label{tab:bin-sim-smry-lg}Averaged evaluation metrics of estimators of area level means across 1,000 binary response simulations for SRS, PPS1, and PPS2 designs for a sample size of one hundred units per area.}
\centering
\fontsize{7}{9}\selectfont
\begin{tabular}[t]{llrrrr}
\toprule
Design & Method & RMSE (x 100) & MAE (x 100) & MIL (x 100) & 90\% Int. Cov.\\
\midrule
 SRS& Hájek & 4.4 & 3.5 & 14.5 & 90\\
\cmidrule{2-6}
 & GREG & 4.2 & 3.3 & 13.8 & 90\\
\cmidrule{2-6}
 & Unwt & 4.1 & 3.3 & 13.4 & 90\\
\cmidrule{2-6}
 & Wt & 4.1 & 3.3 & 13.4 & 90\\
\cmidrule{2-6}
& WtRscl & 4.1 & 3.3 & 13.4 & 90\\
\cmidrule{1-6}
 PPS1 & Hájek & 4.9 & 3.9 & 15.8 & 89\\
\cmidrule{2-6}
 & GREG & 4.7 & 3.7 & 15.0 & 89\\
\cmidrule{2-6}
 & Unwt & 4.6 & 3.6 & 13.4 & 87\\
\cmidrule{2-6}
 & Wt & 4.5 & 3.6 & 13.4 & 87\\
\cmidrule{2-6}
 & WtRscl & 4.5 & 3.6 & 14.6 & 89\\
\cmidrule{1-6}
PPS2 & Hájek & 6.3 & 5.0 & 20.8 & 88\\
\cmidrule{2-6}
 & GREG & 6.2 & 4.9 & 20.2 & 88\\
\cmidrule{2-6}
 & Unwt & 7.1 & 5.6 & 13.3 & 66\\
\cmidrule{2-6}
 & Wt & 5.9 & 4.7 & 13.4 & 75\\
\cmidrule{2-6}
& WtRscl & 5.9 & 4.7 & 19.5 & 89\\
\bottomrule
\end{tabular}
\end{table}

%% file: 2023-09-19_draft.bbl
\begin{thebibliography}{10}

\bibitem{pfeffermann_new_2013}
D.~Pfeffermann, ``New {Important} {Developments} in {Small} {Area}
  {Estimation},'' {\em Statistical Science}, vol.~28, pp.~40--68, Feb. 2013.

\bibitem{rao_small_2015}
J.~N.~K. Rao and I.~Molina, {\em Small {Area} {Estimation}}.
\newblock Wiley {Series} in {Survey} {Methodology}, John Wiley \& Sons,
  2nd~ed., Aug. 2015.

\bibitem{ghosh_small_2020}
M.~Ghosh, ``Small area estimation: its evolution in five decades,'' {\em
  Statistics in Transition}, vol.~21, Sept. 2020.

\bibitem{bell_overview_2016}
W.~R. Bell, W.~W. Basel, and J.~J. Maples, ``An {Overview} of the {U}.{S}.
  {Census} {Bureau}'s {Small} {Area} {Income} and {Poverty} {Estimates}
  {Program},'' in {\em Analysis of {Poverty} {Data} by {Small} {Area}
  {Estimation}}, pp.~349--378, John Wiley \& Sons, Ltd, 2016.

\bibitem{marhuenda_poverty_2017}
Y.~Marhuenda, I.~Molina, D.~Morales, and J.~N.~K. Rao, ``Poverty mapping in
  small areas under a twofold nested error regression model,'' {\em Journal of
  the Royal Statistical Society: Series A (Statistics in Society)}, vol.~180,
  no.~4, pp.~1111--1136, 2017.

\bibitem{corral_pull_2020}
P.~Corral, I.~Molina, and M.~Nguyen, {\em Pull {Your} {Small} {Area}
  {Estimates} {Up} by the {Bootstraps}}.
\newblock Policy {Research} {Working} {Papers}, The World Bank, May 2020.

\bibitem{utazi_geospatial_2020}
C.~E. Utazi, J.~Wagai, O.~Pannell, F.~T. Cutts, D.~A. Rhoda, M.~J. Ferrari,
  B.~Dieng, J.~Oteri, M.~C. Danovaro-Holliday, A.~Adeniran, and A.~J. Tatem,
  ``Geospatial variation in measles vaccine coverage through routine and
  campaign strategies in {Nigeria}: {Analysis} of recent household surveys,''
  {\em Vaccine}, vol.~38, pp.~3062--3071, Mar. 2020.

\bibitem{hogg_two-stage_2023}
J.~Hogg, J.~Cameron, S.~Cramb, P.~Baade, and K.~Mengersen, ``A {Two}-{Stage}
  {Bayesian} {Small} {Area} {Estimation} {Method} for {Proportions},'' Aug.
  2023.
\newblock arXiv:2306.11302 [stat].

\bibitem{erciulescu_model-based_2019}
A.~L. Erciulescu, N.~B. Cruze, and B.~Nandram, ``Model-based county level crop
  estimates incorporating auxiliary sources of information,'' {\em Journal of
  the Royal Statistical Society: Series A (Statistics in Society)}, vol.~182,
  no.~1, pp.~283--303, 2019.

\bibitem{horvitz1952}
D.~G. Horvitz and D.~J. Thompson, ``A generalization of sampling without
  replacement from a finite universe,'' {\em Journal of the American
  Statistical Association}, vol.~47, no.~260, pp.~663--685, 1952.

\bibitem{hajek_discussion_1971}
J.~Hájek, ``Discussion of "{An} essay on the logical foundations of survey
  sampling, part {I}” , by {D}. {Basu}.,'' in {\em Foundations of
  {Statistical} {Inference}} (V.~P. Godambe and D.~A. Sprott, eds.), Toronto:
  Holt, Rinehart and Winston, 1971.

\bibitem{golding_mapping_2017}
N.~Golding, R.~Burstein, J.~Longbottom, A.~J. Browne, N.~Fullman,
  A.~Osgood-Zimmerman, L.~Earl, S.~Bhatt, E.~Cameron, D.~C. Casey,
  L.~Dwyer-Lindgren, T.~H. Farag, A.~D. Flaxman, M.~S. Fraser, P.~W. Gething,
  H.~S. Gibson, N.~Graetz, L.~K. Krause, X.~R. Kulikoff, S.~S. Lim, B.~Mappin,
  C.~Morozoff, R.~C. Reiner, A.~Sligar, D.~L. Smith, H.~Wang, D.~J. Weiss,
  C.~J.~L. Murray, C.~L. Moyes, and S.~I. Hay, ``Mapping under-5 and neonatal
  mortality in {Africa}, 2000–15: a baseline analysis for the {Sustainable}
  {Development} {Goals},'' {\em The Lancet}, vol.~390, pp.~2171--2182, Nov.
  2017.

\bibitem{diggle_model-based_2016}
P.~J. Diggle and E.~Giorgi, ``Model-{Based} {Geostatistics} for {Prevalence}
  {Mapping} in {Low}-{Resource} {Settings},'' {\em Journal of the American
  Statistical Association}, vol.~111, pp.~1096--1120, July 2016.

\bibitem{fay1979}
R.~E. Fay and R.~A. Herriot, ``Estimates of income for small places: An
  application of james-stein procedures to census data,'' {\em Journal of the
  American Statistical Association}, vol.~74, no.~366, pp.~269--277, 1979.

\bibitem{parker_unit_2020}
P.~A. Parker, R.~Janicki, and S.~H. Holan, ``Unit {Level} {Modeling} of
  {Survey} {Data} for {Small} {Area} {Estimation} {Under} {Informative}
  {Sampling}: {A} {Comprehensive} {Overview} with {Extensions},'' {\em
  arXiv:1908.10488 [stat]}, Jan. 2020.
\newblock arXiv: 1908.10488.

\bibitem{binder1983}
D.~A. Binder, ``On the variances of asymptotically normal estimators from
  complex surveys,'' {\em International Statistical Review / Revue
  Internationale de Statistique}, vol.~51, no.~3, pp.~279--292, 1983.

\bibitem{savitsky_bayesian_2016}
T.~D. Savitsky and D.~Toth, ``Bayesian estimation under informative sampling,''
  {\em Electronic Journal of Statistics}, vol.~10, no.~1, pp.~1677--1708, 2016.

\bibitem{pfeffermann1998}
D.~Pfeffermann, C.~J. Skinner, D.~J. Holmes, H.~Goldstein, and J.~Rasbash,
  ``Weighting for unequal selection probabilities in multilevel models,'' {\em
  Journal of the Royal Statistical Society. Series B (Statistical
  Methodology)}, vol.~60, no.~1, pp.~23--40, 1998.

\bibitem{rabe-hesketh2006}
S.~Rabe-Hesketh and A.~Skrondal, ``Multilevel modelling of complex survey
  data,'' {\em Journal of the Royal Statistical Society: Series A (Statistics
  in Society)}, vol.~169, no.~4, pp.~805--827, 2006.

\bibitem{asparouhov2006}
T.~Asparouhov, ``General multi-level modeling with sampling weights,'' {\em
  Communications in Statistics - Theory and Methods}, vol.~35, pp.~439--460, 04
  2006.

\bibitem{slud2020}
E.~V. Slud, ``Model-assisted estimation of mixed-e{ff}ect model parameters in
  complex surveys,'' tech. rep., 2020.

\bibitem{savitsky_pseudo_2022}
T.~D. Savitsky and M.~R. Williams, ``Pseudo {Bayesian} {Mixed} {Models} under
  {Informative} {Sampling},'' {\em Journal of Official Statistics}, vol.~38,
  pp.~901--928, Aug. 2022.

\bibitem{han2021}
Q.~Han and J.~A. Wellner, ``Complex sampling designs: Uniform limit theorems
  and applications,'' {\em The Annals of Statistics}, vol.~49, pp.~459--485, 02
  2021.

\bibitem{williams2021}
M.~R. Williams and T.~D. Savitsky, ``Uncertainty estimation for pseudo-bayesian
  inference under complex sampling,'' {\em International Statistical Review},
  vol.~89, no.~1, pp.~72--107, 2021.

\bibitem{leon-novelo_fully_2019}
L.~G. León-Novelo and T.~D. Savitsky, ``Fully {Bayesian} estimation under
  informative sampling,'' {\em Electronic Journal of Statistics}, vol.~13,
  pp.~1608--1645, Jan. 2019.

\bibitem{parker_computationally_2022}
P.~A. Parker, S.~H. Holan, and R.~Janicki, ``Computationally efficient
  {Bayesian} unit-level models for non-{Gaussian} data under informative
  sampling with application to estimation of health insurance coverage,'' {\em
  The Annals of Applied Statistics}, vol.~16, pp.~887--904, June 2022.

\bibitem{battese1988}
G.~E. Battese, R.~M. Harter, and W.~A. Fuller, ``An error-components model for
  prediction of county crop areas using survey and satellite data,'' {\em
  Journal of the American Statistical Association}, vol.~83, no.~401,
  pp.~28--36, 1988.

\bibitem{hodges_richly_2016}
J.~S. Hodges, {\em Richly {Parameterized} {Linear} {Models}: {Additive}, {Time}
  {Series}, and {Spatial} {Models} {Using} {Random} {Effects}}.
\newblock CRC Press, Apr. 2016.

\bibitem{pfeffermann2007}
D.~Pfeffermann and M.~Sverchkov, ``Small-area estimation under informative
  probability sampling of areas and within the selected areas,'' {\em Journal
  of the American Statistical Association}, vol.~102, pp.~1427--1439, 12 2007.

\bibitem{rubin-bleuer_two-phase_2005}
S.~Rubin-Bleuer and I.~Schiopu-Kratina, ``On the two-phase framework for joint
  model and design-based inference,'' {\em The Annals of Statistics}, vol.~33,
  pp.~2789--2810, Dec. 2005.

\bibitem{you2002}
Y.~You and J.~N.~K. Rao, ``A pseudo-empirical best linear unbiased prediction
  approach to small area estimation using survey weights,'' {\em The Canadian
  Journal of Statistics / La Revue Canadienne de Statistique}, vol.~30, no.~3,
  pp.~431--439, 2002.

\bibitem{torabi_mean_2010}
M.~Torabi and J.~N.~K. Rao, ``Mean squared error estimators of small area means
  using survey weights,'' {\em Canadian Journal of Statistics}, vol.~38, no.~4,
  pp.~598--608, 2010.

\bibitem{skinner1989}
C.~J. Skinner, {\em Domain means, regression and multivariate analysis.},
  pp.~59--87.
\newblock Chichester, UK: Wiley, 1st~ed., 1989.

\bibitem{kish_survey_1965}
L.~Kish, {\em Survey {Sampling}}.
\newblock Wiley, Jan. 1965.

\bibitem{kleijn_bernstein-von-mises_2012}
B.~J.~K. Kleijn and A.~W. v.~d. Vaart, ``The {Bernstein}-{Von}-{Mises} theorem
  under misspecification,'' {\em Electronic Journal of Statistics}, vol.~6,
  pp.~354--381, Jan. 2012.
\newblock Publisher: Institute of Mathematical Statistics and Bernoulli
  Society.

\bibitem{ribatet_bayesian_2012}
M.~Ribatet, D.~Cooley, and A.~C. Davison, ``Bayesian {Inference} from
  {Composite} {Likelihoods}, with an {Application} to {Spatial} {Extremes},''
  {\em Statistica Sinica}, vol.~22, no.~2, pp.~813--845, 2012.

\bibitem{miller_asymptotic_2021}
J.~W. Miller, ``Asymptotic normality, concentration, and coverage of
  generalized posteriors,'' {\em The Journal of Machine Learning Research},
  vol.~22, pp.~168:7598--168:7650, Jan. 2021.

\bibitem{thompson_bayesian_2022}
M.~E. Thompson, J.~Sedransk, J.~Fang, and G.~Y. Yi, ``Bayesian inference for a
  variance component model using pairwise composite likelihood with survey
  data,'' {\em Survey Methodology}, vol.~48, pp.~73--93, June 2022.

\bibitem{preston_rescaled_2009}
J.~Preston, ``Rescaled bootstrap for stratified multistage sampling,'' {\em
  Survey Methodology}, vol.~35, pp.~227--234, Dec. 2009.

\bibitem{rue2017}
H.~Rue, A.~Riebler, S.~H. {Sørbye}, J.~B. Illian, D.~P. Simpson, and F.~K.
  Lindgren, ``Bayesian computing with inla: A review,'' {\em Annual Review of
  Statistics and Its Application}, vol.~4, no.~1, pp.~395--421, 2017.

\bibitem{tille2021}
Y.~Tillé and A.~Matei, {\em sampling: Survey Sampling}, 2021.
\newblock R package version 2.9.

\bibitem{institut_national_de_la_statistique_and_icf_guinea_2019}
{Institut National de la Statistique and ICF}, ``Guinea {Demographic} and
  {Health} {Survey} ({EDS} {V}) 2016-18,'' tech. rep., INS/Guinea and ICF,
  Conakry, Guinea, 2019.

\bibitem{noauthor_database_2022}
``Database of {Global} {Administrative} {Areas} 4.1,'' July 2022.

\bibitem{weiss_global_2018}
D.~J. Weiss, A.~Nelson, H.~S. Gibson, W.~Temperley, S.~Peedell, A.~Lieber,
  M.~Hancher, E.~Poyart, S.~Belchior, N.~Fullman, B.~Mappin, U.~Dalrymple,
  J.~Rozier, T.~C.~D. Lucas, R.~E. Howes, L.~S. Tusting, S.~Y. Kang,
  E.~Cameron, D.~Bisanzio, K.~E. Battle, S.~Bhatt, and P.~W. Gething, ``A
  global map of travel time to cities to assess inequalities in accessibility
  in 2015,'' {\em Nature}, vol.~553, pp.~333--336, Jan. 2018.

\bibitem{worldpop_global_2018}
WorldPop, ``Global 100m {Covariates},'' 2018.
\newblock Type: dataset.

\bibitem{worldpop_global_2020}
WorldPop, ``Global 100m {Population} total adjusted to match the corresponding
  {UNPD} estimate,'' 2020.
\newblock Type: dataset.

\bibitem{worldpop_age_2018}
WorldPop, ``Global 100m {Age}/{Sex} {Structures},'' 2018.
\newblock Type: dataset.

\bibitem{paige2022}
J.~Paige, G.-A. Fuglstad, A.~Riebler, and J.~Wakefield, ``Design- and
  model-based approaches to small-area estimation in a low- and middle-income
  country context: Comparisons and recommendations,'' {\em Journal of Survey
  Statistics and Methodology}, vol.~10, pp.~50--80, 02 2022.

\bibitem{paige_spatial_2022}
J.~Paige, G.-A. Fuglstad, A.~Riebler, and J.~Wakefield, ``Spatial aggregation
  with respect to a population distribution: {Impact} on inference,'' {\em
  Spatial Statistics}, vol.~52, p.~100714, Dec. 2022.

\bibitem{r_core_team_r_2023}
{R Core Team}, {\em R: {A} {Language} and {Environment} for {Statistical}
  {Computing}}.
\newblock Vienna, Austria: R Foundation for Statistical Computing, 2023.

\bibitem{tidyverse}
H.~Wickham, M.~Averick, J.~Bryan, W.~Chang, L.~D. McGowan, R.~François,
  G.~Grolemund, A.~Hayes, L.~Henry, J.~Hester, M.~Kuhn, T.~L. Pedersen,
  E.~Miller, S.~M. Bache, K.~Müller, J.~Ooms, D.~Robinson, D.~P. Seidel,
  V.~Spinu, K.~Takahashi, D.~Vaughan, C.~Wilke, K.~Woo, and H.~Yutani,
  ``Welcome to the {tidyverse},'' {\em Journal of Open Source Software},
  vol.~4, no.~43, p.~1686, 2019.

\bibitem{sf}
E.~Pebesma, ``{Simple Features for R: Standardized Support for Spatial Vector
  Data},'' {\em {The R Journal}}, vol.~10, no.~1, pp.~439--446, 2018.

\bibitem{terra}
R.~J. Hijmans, {\em terra: Spatial Data Analysis}, 2023.
\newblock R package version 1.7-3.

\bibitem{lumley2011}
T.~Lumley, {\em Complex Surveys: A Guide to Analysis Using R}.
\newblock John Wiley \& Sons, 09 2011.

\bibitem{simpson_penalising_2017}
D.~Simpson, H.~Rue, A.~Riebler, T.~G. Martins, and S.~H. Sørbye, ``Penalising
  {Model} {Component} {Complexity}: {A} {Principled}, {Practical} {Approach} to
  {Constructing} {Priors},'' {\em Statistical Science}, vol.~32, pp.~1--28,
  Feb. 2017.

\end{thebibliography}
